\documentclass[12pt]{article}
\usepackage{authblk}
\usepackage{array}
\usepackage{amsmath}
\usepackage{amsthm}
\usepackage{geometry}
\usepackage{mathtools}
\usepackage{commath}
\usepackage{mathrsfs}
\usepackage[mathscr]{eucal}
\usepackage{bm}
\usepackage{esvect}
\usepackage{tikz}
\usepackage{booktabs}
\usepackage{amssymb}
\usepackage{graphics}
\usepackage{picinpar}
\usepackage{float}
\usepackage{color}
\usepackage{esint}
\usepackage{subfigure}
\usepackage{caption}
\usepackage{graphicx}
\usepackage{lipsum}
\usepackage{listings}
\usepackage{lineno}
\usepackage{appendix}

\title{Spectral condensation in quasi-geostrophic turbulence above small-scale topography}
\date{}
\author[1]{Lin-Fan Zhang}
\author[1,2]{Jin-Han Xie\thanks{jinhanxie@pku.edu.cn}}
\affil[1]{\small{\textit{Department of Mechanics and Engineering Science at College of Engineering, State Key Laboratory for Turbulence and Complex Systems, Beijing 100871, P. R. China}}}
\affil[2]{\small{\textit{Joint Laboratory of Marine Hydrodynamics and Ocean Engineering, Laoshan Laboratory, Shandong 266237, P. R. China}}}
\newcommand*{\circled}[1]{\lower .7ex \hbox{\tikz\draw (0pt,0pt)%
		circle (.5em) node {\makebox[1em][c]{\small #1}};}}

\begin{document}
\bibliographystyle{unsrt}
\maketitle
%\linenumbers

\newcommand{\mi}{\mathrm{i}}
\newcommand{\me}{\mathrm{e}}
\newcommand{\mdif}{\mathrm{d}}
\newcommand{\com}[1]{{\color{red}[#1]}}
\newcommand{\rev}[1]{{\textcolor{blue}{#1}}}

\begin{abstract}
    Sea-floor topography is essential for oceanic fluid dynamics in many perspectives, and it is believed to enhance energy dissipation to oceanic flows. 
    This study numerically examines the impact of small-scale topography on the dynamic of quasi-geostrophic barotropic flows and finds that small-amplitude topography enhances upscale energy flux and leads to condensation, which contradicts the common understanding. 
    Topography-induced dissipation only happens when its amplitude is stronger than the first critical value.
    And there exists a second critical topography magnitude, corresponding to a second-order phase transition. 
    When the topography magnitude lies between the two critical values, energy simultaneously transfers to both large and small scales.
    When the topography magnitude exceeds the second critical value, energy only transfers downscale.
    The discovery of counterintuitive topography-enhanced energy flux and the critical phenomenon brings new challenges to topography parameterization in ocean models.

\end{abstract}

%\keywords{}
\maketitle

%\textit{Introduction.}-
%(The importance of topography's influence on geostropic turluence)

Earth's rotation profoundly impacts geostrophic eddies, which possess a large portion of the ocean's kinetic energy and tend to transfer energy towards larger scales \cite{GeostrophicTurbulence}.
However, kinetic energy dissipation in the ocean happens at small viscous scales, thus leading to a paradox \cite{marino_resolving_2015}.
Various mechanisms \cite{hogg_kelvin_2011,zhai_significant_2010,sen_global_2008,arbic_estimates_2009} explain how energy transfers from mesoscale eddies to small-scale unbalanced motions and therefore dissipates at small scales, e.g., the interaction between geostrophic eddies and seafloor is responsible for up to $80\%$ of the wind power input in the Southern Ocean \cite{nikurashin_routes_2013}.

Topography influences the ocean circulation on various temporal and spatial scales \cite{merryfield_topographic_1997,merryfield_eddy_1999,nikurashin_impact_2014,trossman_role_2017}.
It affects the stability of the large-scale flows and changes their associated mixing characteristics \cite{chen_dynamics_2015,brown_effects_2019,lacasce_jet_2019,radko_control_2020}.
Simulations of decaying vortex above irregular roughness find topography stabilizes and dramatically perpetuates coherent vortices to transport heat and nutrients throughout the ocean \cite{gulliver_topographic_2022}.
Besides, topography generates internal waves that radiate energy through the domain \cite{klymak_nonpropagating_2018}. 
Topography is also essential in explaining the direct cascade of kinetic energy in oceans \cite{merryfield_ocean_2005}.
Numerical and analytical studies point out that topography catalyzes energy cascade from large-scales of motion to sub-mesoscale eddies and leads to dissipation at the microscale \cite{dewar_topographic_2010,dewar_submesoscale_2011}.
Although the importance of topography is widely acknowledged, the mechanism of geostrophic turbulence and topography interaction is complicated \cite{radko_spin_down_2022b}.
Much attention focused on the bottom pressure torque \cite{hughes_why_2001,olbers_dynamical_2004}, the lee-wave-induced drag \cite{arbic_connecting_2019,eden_closure_2021,klymak_parameterizing_2021} and the topographic steering of low Rossby number flows \cite{marshall_topographic_1995,wahlin_topographic_2002}.
Based on equilibrium statistical mechanics, the interaction between topography and eddies generates a secondary circulation that can accelerate mean flows, referred to as the Neptune effect \cite{holloway_systematic_1987}. 

Although high-performance computers are greatly improved, roughness-resolving global models for climate simulations are still too expensive \cite{radko_generalized_2023}, and the parameterization of the topographic effect remains an open question.
Under the assumption that flows exhibit a propensity to optimize statistical entropy, Holloway \cite{holloway_representing_1992} parameterizes the Neptune effect by a propensity for along-topography flow \cite{holloway_representing_2009}.
Considering the scale separation between mean flows and rough topography, Vanneste \cite{vanneste_enhanced_2000} reduced the small-scale topographic effect to an additional Ekman friction that dissipates the energy of large-scale flows.
Radko \cite{radko_topographic_2017,radko_spin_down_2022a,radko_spin_down_2022b} parameterizes the small-scale topography as topographically induced drag that substantially slows the large-scale motion. 
In this letter, we present a counterintuitive topography-enhanced energy upscales flux that leads to spectral condensation.

The quasi-geostrophic (QG) equation, capturing the material invariant evolution of potential vorticity, is a suitable model to describe the synoptic scales or weather scales in the ocean \cite{vallis_essentials_2019}.
This letter considers the homogeneous QG model in single-layer shallow water with irregular bottom roughness on an f-plane (cf.\cite{pedlosky_geophysical_1987,bretherton_two-dimensional_1976}).
Considering scales much smaller than the deformation radius $L_d$, the material invariant potential vorticity reduces to $q=Q + h$, where $Q=\nabla^2\psi$ is the vorticity with $\psi$ the streamfunction, and the topography $h$ is the local variation of layer thickness normalized by the Rossby number.
The QG system conserves energy $\mathcal{E}$ and potential enstrophy $\mathcal{Z}$: 
\begin{equation}\label{eq:conservative}
	\mathcal{E}=\int |\nabla\psi|^2 \mdif \boldsymbol{x},
	\mathcal{Z}=\int (Q+h)^2 \mdif \boldsymbol{x}.
\end{equation}

Including the forcing and dissipation effects, the dimensionless QG equation becomes
\begin{equation}\label{eq:vorticity}
	\partial_t Q + J(\psi,Q) + J(\psi, h) = -\alpha Q + \nu\nabla^6 Q + F,
\end{equation}
where $J(a,b)=a_x b_y - a_y b_x$.
The first linear damping term on the right-hand side captures the effects of top and bottom frictions; the second hyper-viscous term dissipates downscale energy flux and $F$ denotes the external forcing which we specify below.

%\textit{Numerical scheme and parameters.}- 
We perform numerical simulations using a Fourier pseudo-spectral method with $2/3$ dealiasing in space.
The resolution is $512^2$ in a domain size of $(2\pi)^2$.
Temporally, we apply a fourth-order explicit Runge-Kutta scheme where the linear terms are solved by an integrating-factor method, and the nonlinear terms are explicitly approximated \cite{cox_exponential_2002}. 
Forcing has the form $F=M_F k_f^{1/2}\mathsf{F}$.
Here, $\mathsf{F}$ is white-noise in time, isotropic in space, and centers around wavenumber $|\boldsymbol{k}| = k_f$, and with the correlation
\begin{equation}
	\left\langle\mathsf{F}(\boldsymbol{x}_1,t_1)\mathsf{F}(\boldsymbol{x}_2,t_2)\right\rangle={J}_0(k_\mathrm{f}|\boldsymbol{x_1}-\boldsymbol{x_2}|)\delta(t_1-t_2),
\end{equation}
where $\left\langle \cdot \right\rangle$ is an ensemble average, ${J}_0$ is the zeroth-order Bessel function, and $\delta$ is the Dirac function.
The forcing magnitude $M_F$ controls the energy injection rate $\varepsilon\sim M_F^2/2$.

We define topography, $h(x,y)=M_h R(x,y)$, as a time-independent spatially random field centers around a topography wavenumber $k_h$.
$M_h$ controls the topography magnitude and $max[R]=1$.
The form of topography $R$ is described as:
\begin{equation}
	{R}(x,y)=\mathcal{F}^{-1}\left\{A\me^{-\frac{(|\boldsymbol{k}|-k_h)^2}{2\delta_k^2}}\right\},
\end{equation}
where $\mathcal{F}^{-1}$ denotes the inverse Fourier transform, $A$ is a complex field with uniformly distributed phases in the range of $[0,2\pi]$, and $\delta_k$ captures the width of the topography in the spectral space.
We provide an illustration of topography in the Supplemental Material Section 1.

When the bottom is flat, i.e., $h=0$, (\ref{eq:vorticity}) reduces to the forced-dissipative two-dimensional vorticity equation, known for its energy inverse cascade \cite{kraichnan_inertial_1967,kraichnan_two_dimensional_1980}.
Consequently, the linear damping term establishes a damping scale $L_\alpha\approx\alpha^{-3/2} \varepsilon^{1/2}$.
When $L_\alpha$ is larger than the domain size, energy condensates at the domain size with coherent vortices \cite{smith_bose_1993,chertkov_dynamics_2007}.
To prevent energy condensation in the absence of topography, we choose $L_\alpha$ smaller than the domain size.

Based on the multi-scale analysis \cite{radko_spin_down_2022a}, the leading order balance of (\ref{eq:vorticity}) reads $\nabla^2\psi_1 + h =0$. 
Further assuming that the forcing-induced energy injection balances with the topography-induced energy flux at the topographic scale, i.e., $\varepsilon \sim \psi_1 J(\psi_1,h)$, we obtain a normalized topography magnitude $\mathcal{H} = M_h\varepsilon^{-1/3}k_h^{-2/3}$. 

In this work, we consider three energy injection rates  $\varepsilon_{1}=0.002,\varepsilon_{2}=0.008,\varepsilon_{3}=0.032$. 
Meanwhile, we keep the damping scale $L_\alpha$ constant in different cases by justifying the damping coefficient $\alpha$.
We fix the forcing wavenumber $k_{f}=32$ and vary the topographic wavenumber $k_{h,1}=48,k_{h,2}=64$ and $k_{h,3}=96$ in our simulations. 
Thus, we have nine groups of simulations, whose parameters are shown in Table~\ref{tab:param}.  

\begin{table}[H]
%\begin{ruledtabular}
\centering
\begin{tabular}{ccccc}
\toprule
$k_h$& $M_h$& $\alpha$&$\varepsilon$&$\mathcal{H}$\\
%\colrule
\midrule
48 & 0-256 & 0.0063 & $2\times10^{-3}$ & 0-154\\
64 & 0-256 & 0.0063 & $2\times10^{-3}$ & 0-127\\
96 & 0-256 & 0.0063 & $2\times10^{-3}$ & 0-96\\
48 & 0-512 & 0.01 & $8\times10^{-3}$ & 0-194\\
64 & 0-512 & 0.01 & $8\times10^{-3}$ & 0-160\\
96 & 0-512 & 0.01 & $8\times10^{-3}$ & 0-122\\
48 & 0-768 & 0.0159 & $3.2\times10^{-2}$ & 0-183\\
64 & 0-768 & 0.0159 & $3.2\times10^{-2}$ & 0-151\\
96 & 0-768 & 0.0159 & $3.2\times10^{-2}$ & 0-115\\
\bottomrule
\end{tabular}
%\end{ruledtabular}
\caption{\label{tab:param}%
	Parameters used in the QG turbulence simulations.
}
\end{table}

%\textit{Numerical results}- 
We focus on statistics of statistically steady states of the energy and potential enstrophy.
We introduce a filter to present the cross-scale energy flux and show the large-scale structures.
For a field $\phi$, the filtered field $\bar{\phi}$ is defined as
\begin{equation}
	\bar\phi = \mathcal{F}^{-1}\left\{ \sum_{|\boldsymbol{k}|\leq k_l}^{} \hat\phi(\boldsymbol{k}) \right\} \quad \mathrm{with} \quad \hat\phi = \mathcal{F}\left\{\phi\right\}.
\end{equation}
Throughout this letter, we choose $k_l=16$, which is located at the energy inertial range.
Thus, a field $\phi = \overline{\phi} +\phi'$ is decomposed into the slow and fast parts, $\overline{\phi}$ and $\phi'$,  respectively.

\begin{figure}[h]
    \centering
    \includegraphics[width=0.75\linewidth]{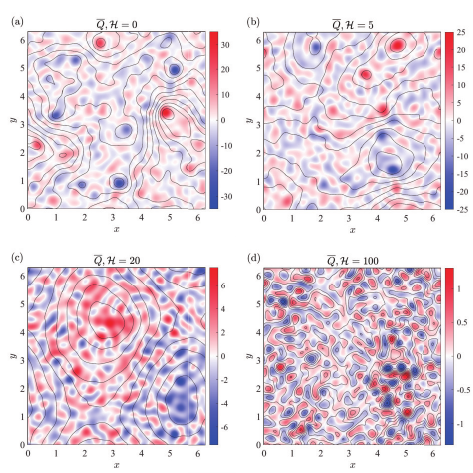}
    \caption{
    Snapshots of statistically steady states of filtered vorticity $\overline{Q}$ with $\mathcal{H}=0,5,20,100$.The black curves are ten contours of the filtered streamfunction $\bar{\psi}$ ranging in $[\min{\bar{\psi}},\max{\bar{\psi}}]$.
    }
    \label{fig:snapshots}
\end{figure}

Figure~\ref{fig:snapshots} displays the snapshots of filtered vorticity $\overline{Q}$ at statistically steady states with energy injection rate $\epsilon_2$, topography scale $k_{h,2}$ and varying topography magnitudes.
Without topography, i.e., $\mathcal{H}=0$, the field of $\overline{Q}$ presents several vortices without the feature of energy condensation.
When the topography magnitude $\mathcal{H}$ increases, the snapshots of $\mathcal{H}=0,5,20$ show that the scale of the dominant vortex grows, and gradually a condensation at the domain size appears.
However, then as the topography magnitude increases, the condensation disappears. E.g., when $\mathcal{H}=100$, structures with scales larger than $1/k_l$ are hard to observe.
More snapshots of $\overline{Q}$ with different $\mathcal{H}$ are presented in Supplemental Material Section 2.
We also quantify the energy condensation using a characteristic length scale in Supplemental Material Section 3.

In the spectral space, we define the energy spectrum
\begin{equation}
    E(K)=\sum_{|\boldsymbol{k}|=K-1/2}^{|\boldsymbol{k}|=K+1/2}\frac{1}{2}|\boldsymbol{k}|^2\hat{\psi}(\boldsymbol{k})\hat{\psi}^*(\boldsymbol{k})
\end{equation}
and the spectral energy flux of QG flow
\begin{equation}
    \Pi(K)=-\sum_{|\boldsymbol{k}|=0}^{K} \frac{1}{2}\hat{\psi}^*\hat{J}(\psi,Q+h)+c.c..
\end{equation}
The energy transfer of the advective and topographic components are
\begin{equation}
T_{adv}(K) = -\sum_{|\boldsymbol{k}|=K-1/2}^{|\boldsymbol{k}|=K+1/2}\frac{1}{2}\hat{\psi}^* \hat{J}(\psi,Q) + c.c, 
\end{equation}
and
\begin{equation}
T_{topo}(K) = -\sum_{|\boldsymbol{k}|=K-1/2}^{|\boldsymbol{k}|=K+1/2}\frac{1}{2}\hat{\psi}^* \hat{J}(\psi,h) + c.c.
\end{equation}
Here, $^*$ and $c.c.$ denote the complex conjugate. $T = T_{adv} + T_{topo}$ is the total energy transfer.

Since the hyper-viscosity dissipation is negligible at large scales, the energy balance at the system scale ($K=1$) is approximately
\begin{equation}\label{eq:linear}
	{T}(K=1) \approx -\alpha {E}(K=1),
\end{equation}
so the total energy transfer at $K=1$ is linear to the energy condensed at the domain size.

\begin{figure}[h]
    \centering
    \includegraphics[width=0.8\linewidth]{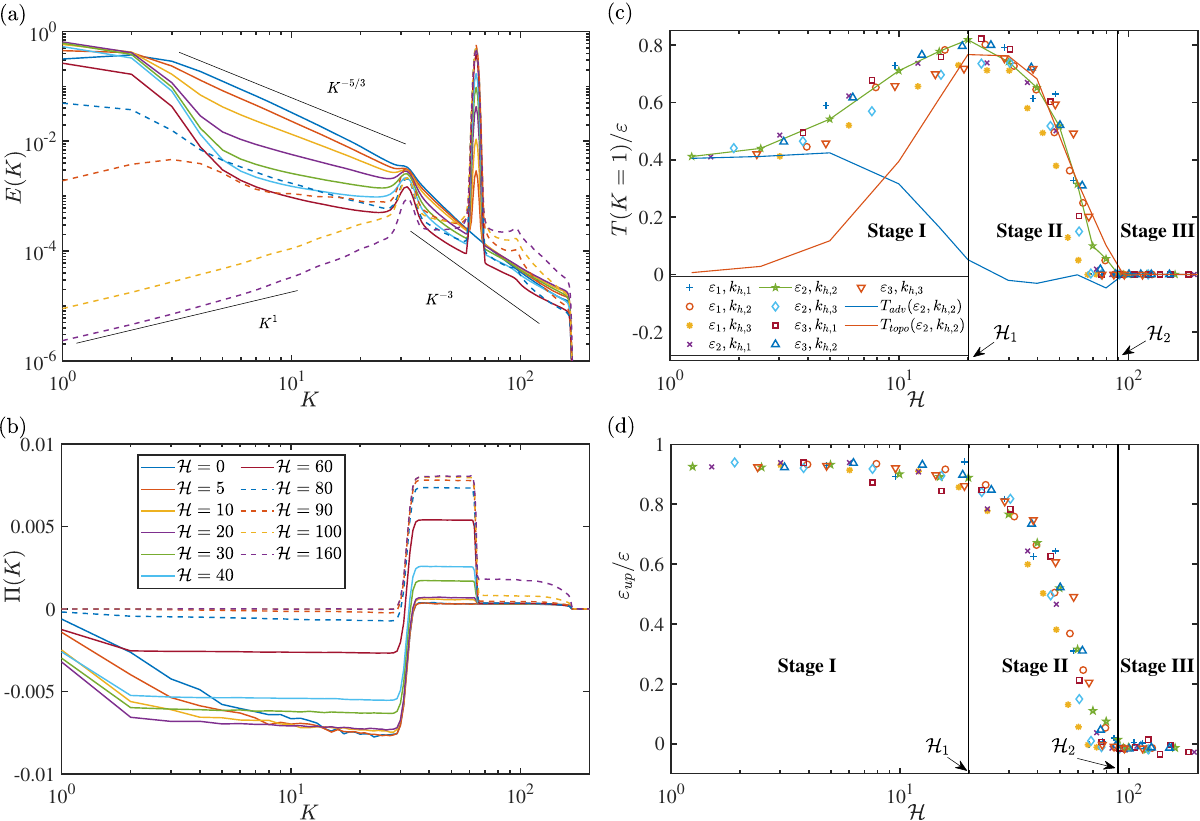}
    \caption{
    The energy spectra (a) and energy fluxes (b) at statistically steady states with $\mathcal{H}$ ranging from $0$ to $160$.
    (a) (b) share the parameters as  $\varepsilon=0.008$ and $k_h=64$ and the same legend.
    (c) shows different components of energy transfer with the normalised topography magnitude with nine combinations of energy injection rates and topography scales.
    (d) shows the dependence of the upscale energy ratio on $\mathcal{H}$.
    (c) and (d) share the same legend.
    Here, $\varepsilon_{1} =0.002,\varepsilon_{2} =0.008,\varepsilon_{3} =0.032$,
    $k_{h,1}=48, k_{h,2}=64, k_{h,3}=96$.
    $\mathcal{H}_1$ and  $\mathcal{H}_2$ denote two critical points.}
    \label{fig:ThreeStages}
\end{figure}

When there is no topography, i.e., $\mathcal{H}=0$, figure~\ref{fig:ThreeStages} (a) shows that at wavenumbers smaller than the forcing wavenumber $k_f$ the energy spectrum follows a Kolmogorov scaling $K^{-5/3}$ of constant energy flux and the energy spectrum larger than $k_f$ follows the scaling $K^{-3}$ of constant enstrophy flux \cite{kraichnan_inertial_1967}.
Meanwhile, the upscale energy flux without topography is not constant due to the linear damping, as shown in figure~\ref{fig:ThreeStages} (b).
As $\mathcal{H}$ increases from $0$, the topographic impacted QG turbulence presents three stages, with two critical values are $\mathcal{H}_1\approx 20$, $\mathcal{H}_2 \approx 90$.
	
In Stage I, i.e. $\mathcal{H}\in(0,\mathcal{H}_1)$, we observe an interesting phenomenon that the peak of the energy spectrum moves to the domain size, as depicted in figure~\ref{fig:ThreeStages}(a).
Concurrently, the energy at mid-scales diminishes, leading to the distinctive feature of energy condensation \cite{smith_bose_1993,xia_spectrally_2009}.
Figure~\ref{fig:ThreeStages}(b) illustrates an increase in energy fluxes at small wavenumbers, expanding the region of constant flux from $K\in[10, 32]$ to $K\in[2,32]$.
Weak topography induces more energy transferred to the system scale, resulting in spectral condensation through the non-local energy transfer, whose details are presented in Supplemental Material Section 4. 
Given that the total energy transfer is directly proportional to the energy condensed at the system scale (cf.(\ref{eq:linear})), $T(K=1)$ in figure~\ref{fig:ThreeStages}(c) reaches its maximum at $\mathcal{H}_1$, corresponding to an energy accumulation at wavenumber $K=1$.
Simultaneously, the topography-induced energy transfer $T_{topo}(K=1)$ increases from $0$ to its peak value, while the advection-induced energy transfer $T_{adv}(K=1)$ monotonously decreases to zero.
To capture the change in upscale energy flux, we introduce the upscale energy flux ratio, $\varepsilon_{up}/\varepsilon$, where $\varepsilon_{up}$ is the upscale energy flux just below the forcing wavenumber.
In Stage I, figure~\ref{fig:ThreeStages} (d) shows that the upscale energy flux ratio remains close to 1, implying the upscale energy flux scenario.
Here, a minor downscale flux is an effect of finite viscosity in numerical simulations.

In Stage II, $\mathcal{H}\in(\mathcal{H}_1,\mathcal{H}_2)$, the energy condensed at the system scale weakens and energy at large scales undergoes gradual damping due to the interaction between topography and the flow as shown in figure~\ref{fig:ThreeStages}(a). 
Simultaneously, as shown in figure~\ref{fig:ThreeStages}(b) and (d), the upscale energy ratio steadily decreases, indicating the emergence of a dual-energy flux at $\mathcal{H}_1$ and a transition in the energy dynamics.
This behaviour aligns with the common understanding of topography-induced extra damping \cite{vanneste_enhanced_2000,radko_topographic_2017,radko_spin_down_2022a}.
%Topography prompts more energy transferred downscale and damped at the topographic wavenumber $k_h$. 
Figure~\ref{fig:ThreeStages}(c) reveals that during Stage $2$, $T(K=1)$ gradually drops to zero as the normalized magnitude of topography $\mathcal{H}$ approaches the second critical point $\mathcal{H}_2$ and the dominant topography-related energy transfer $T_{topo}$ balances the injected energy rate $\epsilon$.
Thus, figure~\ref{fig:ThreeStages}(c) and (d) show that with the normalization $\mathcal{H}=M_h \varepsilon^{-1/3}k_h^{-2/3}$, results from various sets of simulations with differing energy injection rates and topographic scales converge, which confirms the dominant mechanism of energy balance. 

In Stage III when $\mathcal{H}$ exceeds $\mathcal{H}_2$, Figure~\ref{fig:ThreeStages}(a) shows that the energy spectra with $k<k_f$ abruptly change from the $K^{-5/3}$ scaling to the quasi-equilibrium scaling $K^{1}$ \cite{kraichnan_two_dimensional_1980,dallas_statistical_2015,gorce_statistical_2022}.
Figure~\ref{fig:ThreeStages}(b) and (d) show that the upscale energy ratio approaches zero, and all energy transfers downscale.
Concurrently, figure~\ref{fig:ThreeStages}(c) illustrates that $T(K=1)$, $T_{adv}(K=1)$ and $T_{topo}(K=1)$ are both very weak.

%% The filtered equation
%%%%%%%%%%%%%%%%%%%%%%%%%%%%%%%%%

Applying the filter to (\ref{eq:vorticity}), we obtain 
\begin{equation}\label{eq_meanQ}
    \frac{\partial \overline{Q}}{\partial t}+J(\overline{\psi},\overline{Q})+ \Omega_l = -\alpha \overline{Q} + \nu\nabla^6 \overline{Q},
\end{equation}
where $\Omega_l = \overline{J(\psi,Q+h)} - J(\overline{\psi},\overline{Q})$.
Since the forcing centers around wavenumber $k_f>k_l$, we have $\bar{F}=0$.
Multiplying $-\overline{\psi}$ with (\ref{eq_meanQ}), we obtain the filtered energy equation
\begin{equation}\label{eq:pi}
	\frac{\partial \overline{E}}{\partial t}+ \mathcal{J}+ \Pi_l  = \mathcal{\eta},
\end{equation}
where $ \overline{E} = |\nabla\overline{\psi}|^2/2$ is the large-scale energy and $\mathcal{\eta} = \alpha\overline{\psi}\nabla^2\overline{\psi} -\nu\overline{\psi}\nabla^8\overline{\psi}$ is the dissipation. 
$\mathcal{J}=-\nabla\cdot(\overline{\psi}{\partial_t\nabla\overline{\psi}}) -\overline{\psi}J(\overline{\psi},\overline{Q}) $ is the spatial transport of large-scale kinetic energy. 
$\Pi_l =-\overline{\psi} \Omega_l$ is the sub-filter-scale (SFS) flux. 

\begin{figure}[h]
    \centering    
    \includegraphics[width=0.55\linewidth]{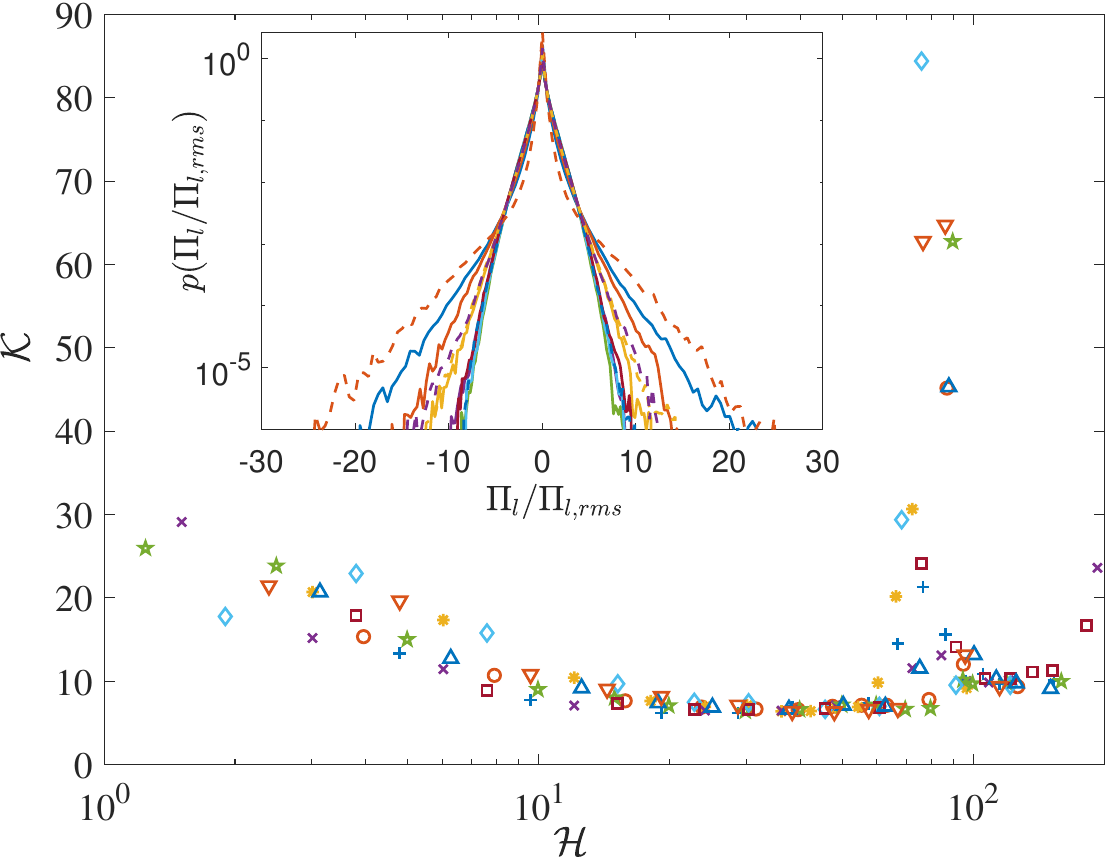}
    \caption{
    The normalised flatness of SFS flux $\Pi_{l}$ with normalised topography magnitude $\mathcal{H}$.
    The inset figure shows the pdfs of the normalised SFS flux $\Pi_{l}/\Pi_{l,rms}$ with $\varepsilon = 0.008$ and $k_h =64$. The legend of the inset figure is the same with figure~\ref{fig:ThreeStages}(a) and (b). The legend of normalised flatness is the same as figure~\ref{fig:ThreeStages}(c) and (d).
    The kurtosis of $\epsilon_1$, $\epsilon_2$, $\epsilon_3$ at $\mathcal{H}=0$ are $28.65$,$29.08$,$30.99$, respectively.
    }
    \label{fig:energyTransfer}
\end{figure}

We calculate SFS flux with different magnitudes of small-scale topography to examine its impact on cross-scale energy flux. 
The inset of figure~\ref{fig:energyTransfer} shows the probability distribution function (PDF) of the SFS flux normalized by its root-mean-square $\Pi_l/\Pi_{l,rms}$ at statistically steady states.
Both upscale and downscale energy transfers in physical space exist simultaneously, and the PDFs of $\Pi_l$ are asymmetric and non-Gaussian, which is similar to the two-dimensional homogeneous isotropic turbulence in \cite{chen_physical_2006,boffetta_energy_2007}.
We define the kurtosis of SFS flux $\Pi_l$ as
\begin{equation}\label{eq:kurtosis}
	\mathcal{K}  = \frac{\left\langle(\Pi_{l}-\left\langle\Pi_l\right\rangle)^4\right\rangle}{\left\langle(\Pi_{l}-\left\langle\Pi_l\right\rangle)^2\right\rangle^2}.
\end{equation}
Fig.~\ref{fig:energyTransfer} shows that the kurtosises with different energy injection rates and topography wavenumber $k_h$ share a universal behavior, confirming an energy-flux-based understanding of the topography-induced system.
When the topography magnitude $\mathcal{H}$ increases from $0$, the kurtosis of SFS flux PDF decreases first.
As the energy transfer to system scale $T(K=1)$ increases to its maximum (cf. figure~\ref{fig:ThreeStages} (c)), kurtosis reaches its minimum around $\mathcal{H}_1\approx 20$.
Then, the kurtosis changes slightly. 
Notably, around the second critical point $\mathcal{H}_2\approx 90$, a prominent peak is observed, indicating a phase transition \cite{vovchenko_non_gaussian_2016}. 
Details of the $\Pi_l$ fields can be found in Supplemental Materials Section 3.

%%%%%%%%%%%%%%%%%%%%%%%%%%%%%%%%%%

%%%%%%%%%%%Discussion%%%%%%%%%%%%
In summary, we discover a novel phenomenon where small-scale topography induces energy condensation in quasi-geostrophic turbulence and find that topography induces phase transition.
The condensation appears with weak topographies and energy transfers upscale.
For an intermediate topography, bidirectional energy flux is observed.
Strong topography eliminates all upscale energy fluxes, which happens beyond a critical value. 
However, the current ocean parameterizations (cf. \cite{holloway_representing_1992,vanneste_enhanced_2000,holloway_representing_2009,nikurashin_routes_2013,klymak_parameterizing_2021,radko_generalized_2023}) are inadequate in capturing this topography-enhanced energy flux presented in this letter, which challenges ocean modelling.
This energy condensation is similar to the large-scale coherent structures observed in two-dimensional active matter turbulence \cite{linkmann_phase_2019}, 
and the topography-enhanced downscale geostrophic energy flux resembles the effect of near-inertial waves \cite{xie_downscale_2020}, but the link between the mechanisms behind these systems remains to be explored.

\textit{Acknowledgement} This project has received financial support from the National Natural Science Foundation of China (NSFC) under grant NO. 92052102 and 12272006, and from the Laoshan Laboratory under grant NO. 2022QNLM010201.

\clearpage

\appendix

\begin{center}
    \large{Supplemental Meterials for `` Spectral condensation in quasi-geostrophic turbulence above small-scale topography''}\\
    \normalsize{Lin-Fan Zhang, Jin-Han Xie}
\end{center}

In the Supplemental Materials, we provide the following insights:
In Section \ref{topo}, we introduce the small-scale topography.
Section \ref{snapshots} includes snapshots of vorticity fields, filtered vorticity fields, and SFS flux fields.
Section \ref{scale} quantifies energy condensation with a characteristic length scale.
In Section \ref{transfer}, we illustrate the energy transfer between two modes.

\section{Topography setting}\label{topo}
\begin{figure}[H]
	\centering
    \includegraphics[width=0.55\linewidth]{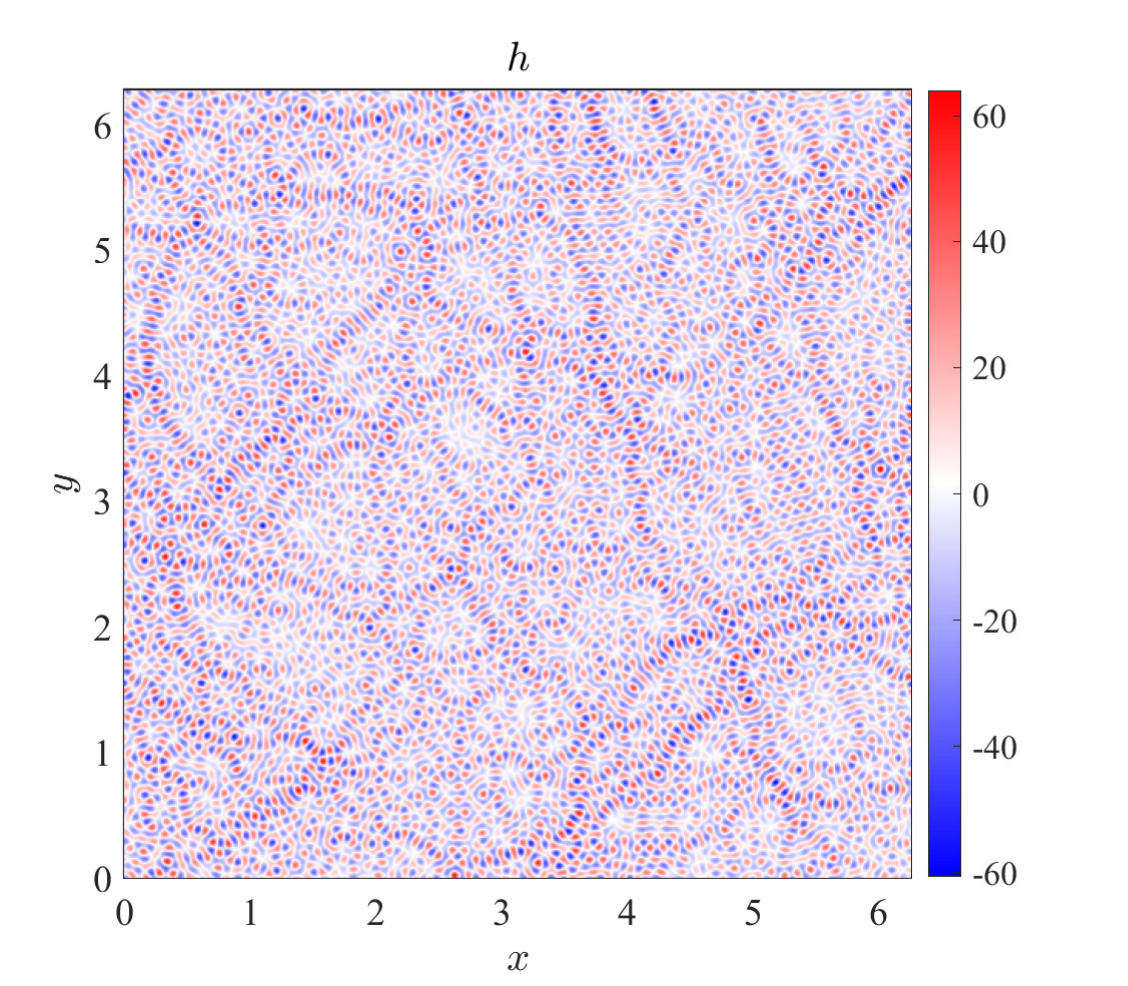}
	\caption{The field of topography with $M_h=64$, $k_h=64$ and $\delta_k=1$.}
	\label{fig:topography}
\end{figure}

We define topography as $h(x,y)=M_h R(x,y)$ at a topography wavenumber $k_h$ with the width $\delta_k$.
Figure~\ref{fig:topography} gives an example of topography $h$ with $M_h=64$,$k_h=64$,$\delta_k=1$.

\section{Snapshots of flitered vorticity and SFS flux}\label{snapshots}
\begin{figure}
	\centering
     \includegraphics[width=0.30\linewidth]{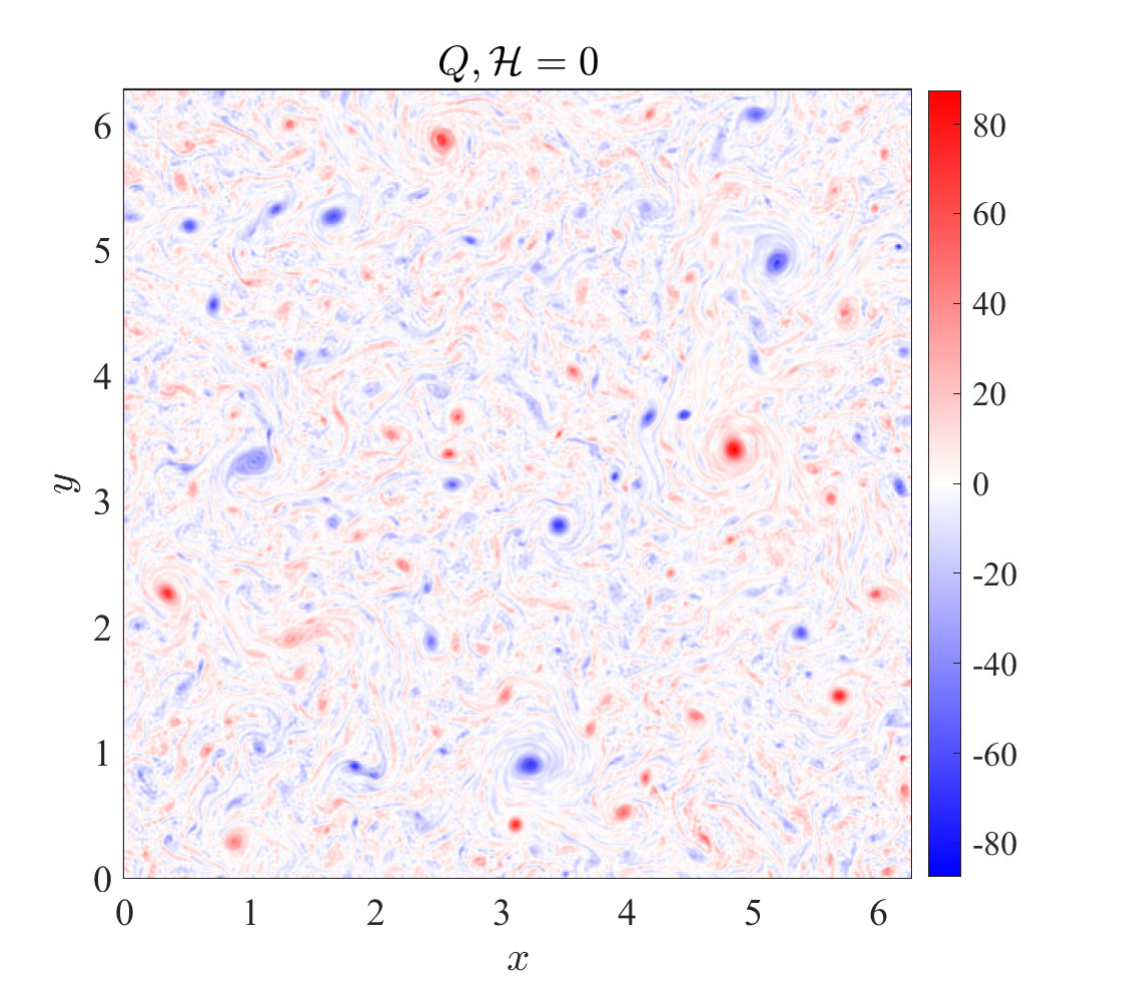}
    \includegraphics[width=0.30\linewidth]{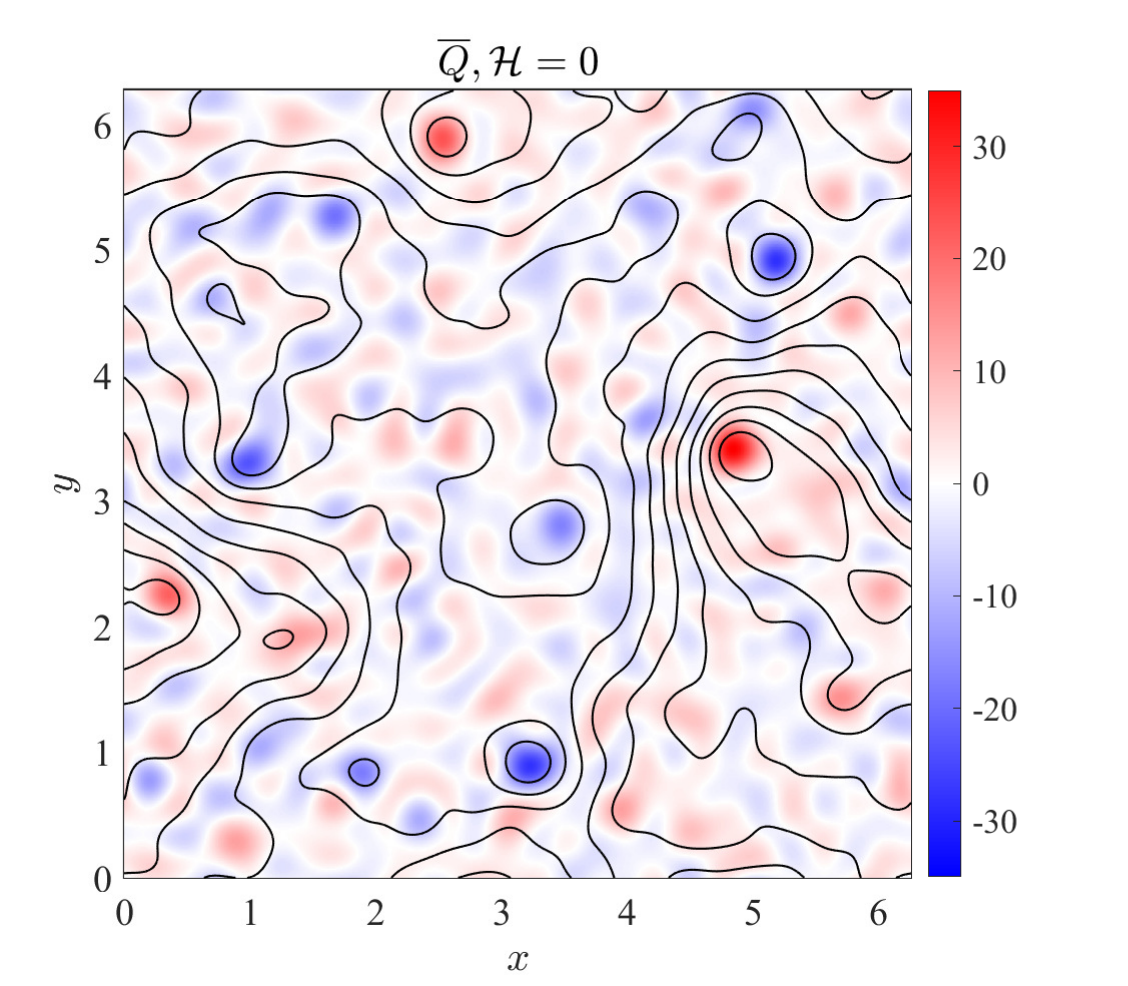}
    \includegraphics[width=0.30\linewidth]{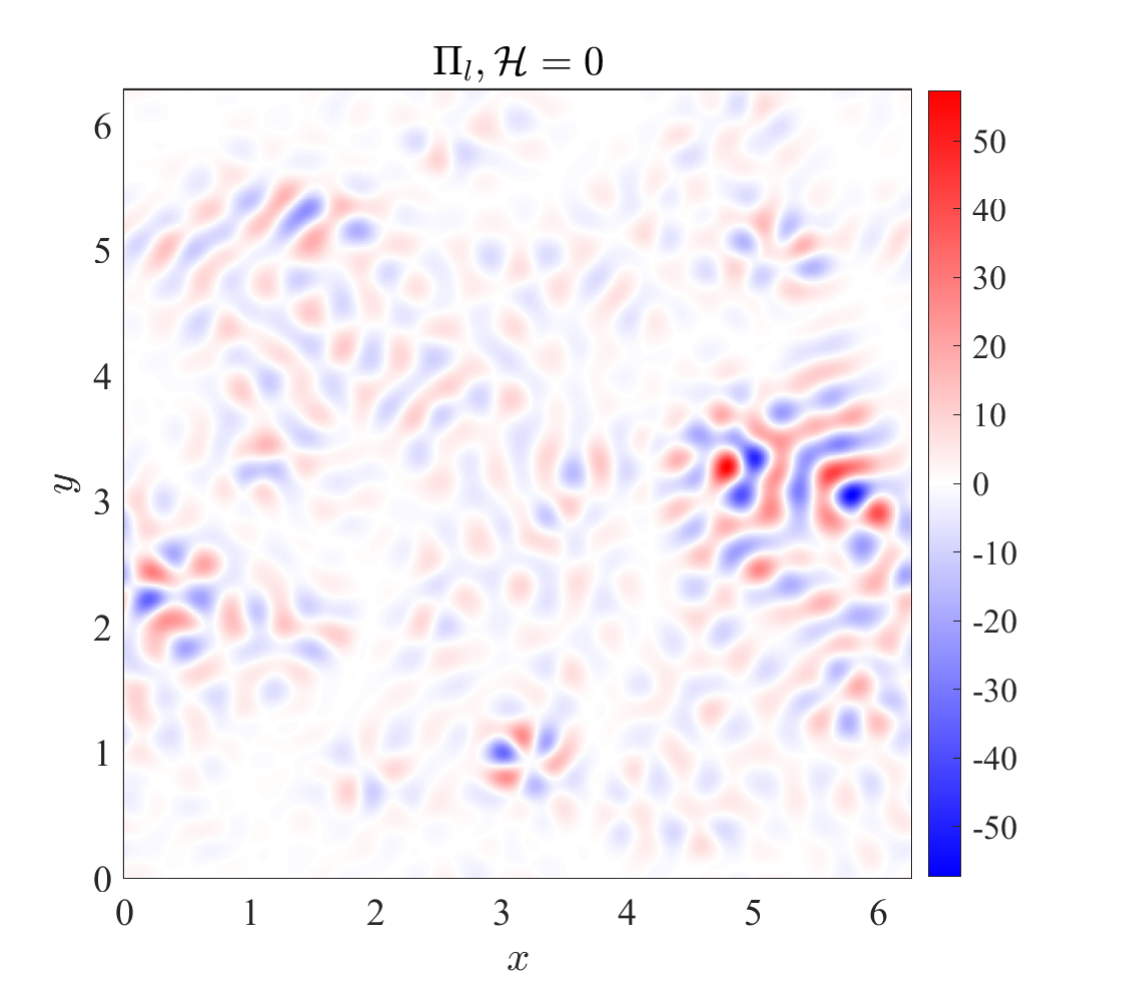}
	\caption{The snapshots of $Q$, $\overline{Q}$, $\Pi_l$ at $\mathcal{H}=0$ with $\epsilon=0.008,k_h=64$. The black curves in the snapshots of $\bar{Q}$ are the contours of the filtered stream-function $\bar{\psi}$.
	Ten contours range in $[-1.34,0.65]$.}
	\label{fig:Mh0SnapShots}
\end{figure}

We provide snapshots of vorticity $Q$, filtered vorticity $\overline{Q}$, and SFS flux $\Pi_l$ without topography at the statistically steady state in Figure~\ref{fig:Mh0SnapShots}. In these fields, no energy condensation is observed.

\begin{figure}
	\centering
	\includegraphics[width=0.30\linewidth]{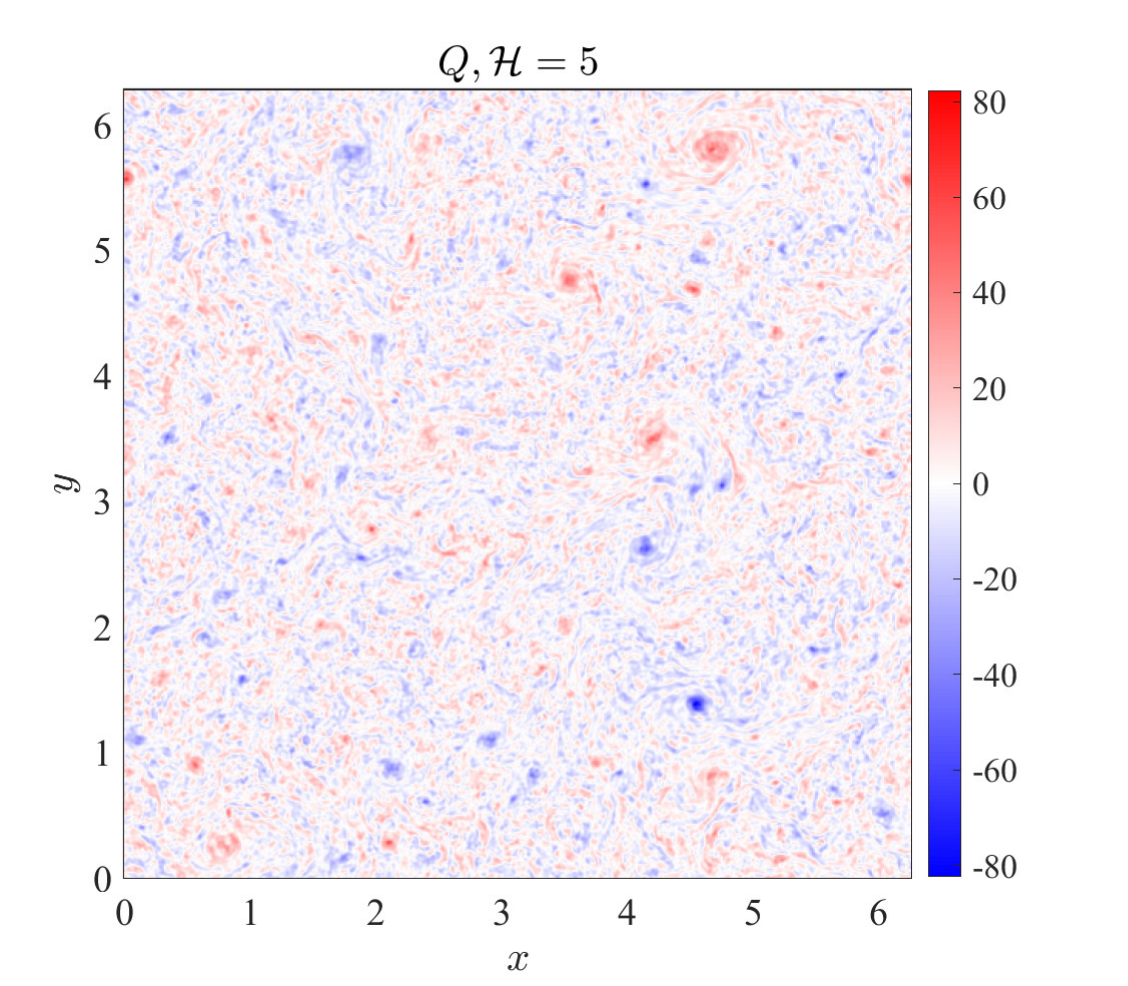}
    \includegraphics[width=0.30\linewidth]{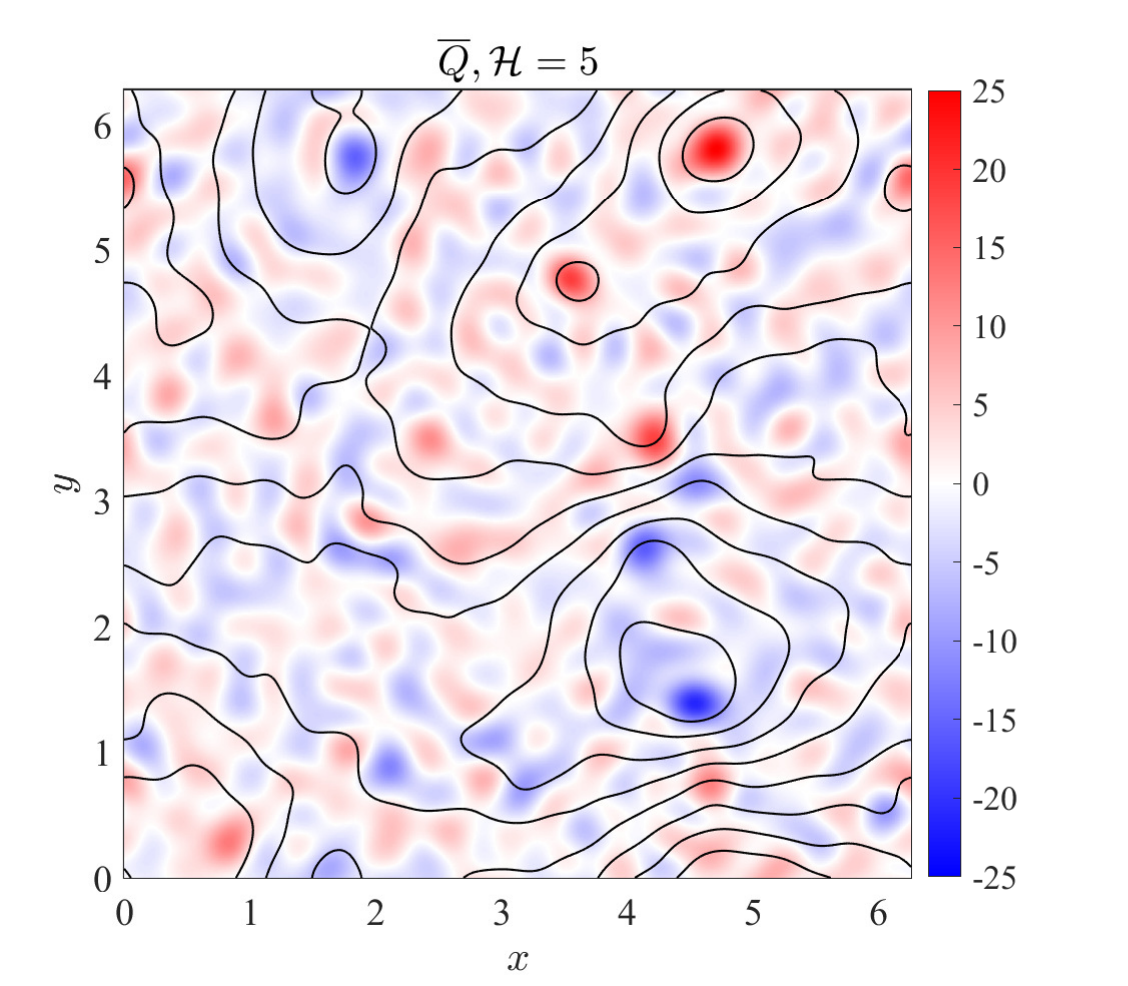}
    \includegraphics[width=0.30\linewidth]{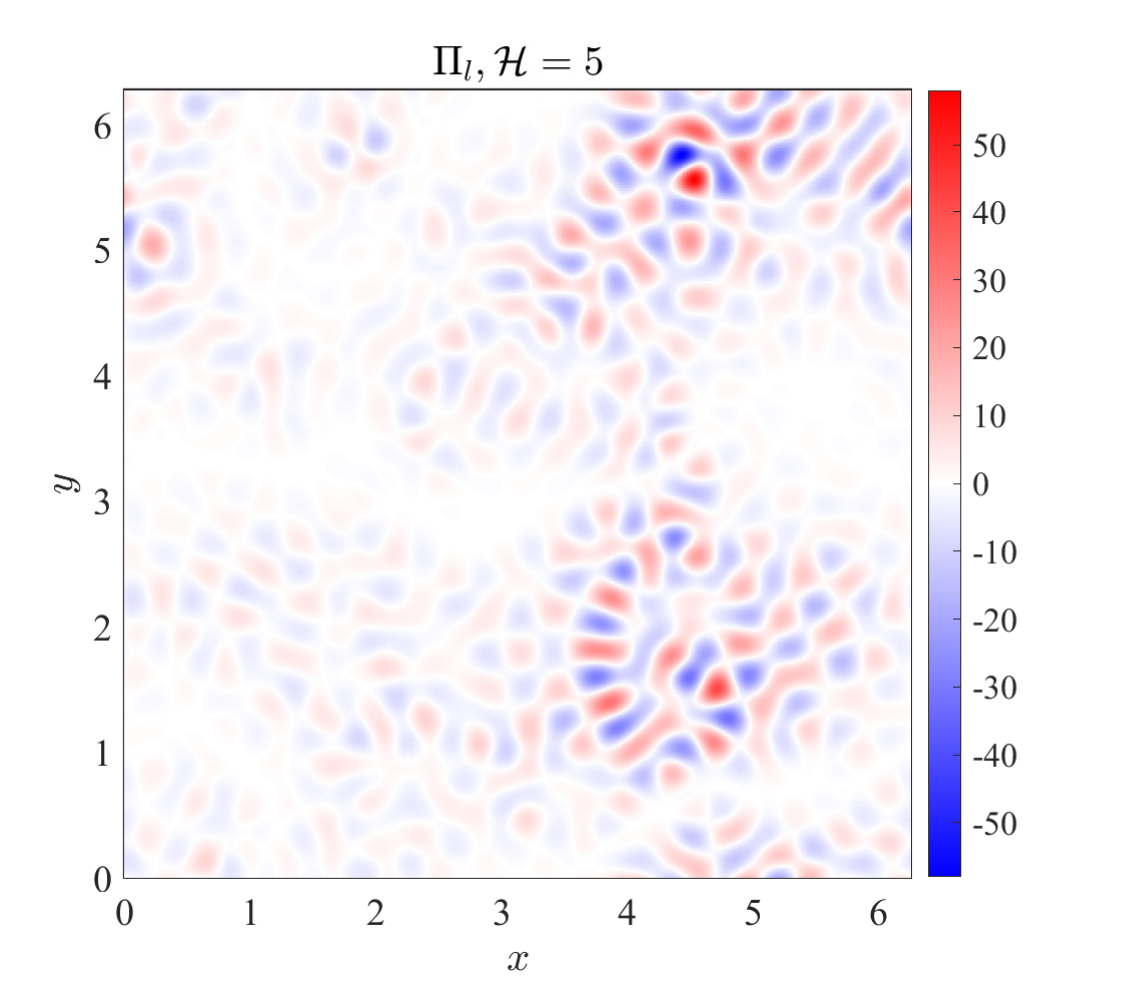}
	\caption{The snapshots of $Q$, $\overline{Q}$, $\Pi_l$ at $\mathcal{H}=5$ with $\epsilon=0.008,k_h=64$. The black curves in the snapshots of $\bar{Q}$ are the contours of the filtered stream-function $\bar{\psi}$.
	Ten contours range in $[-1.41,1.33]$.}
	\label{fig:Mh16SnapShots}
\end{figure}

\begin{figure}
	\centering
	\includegraphics[width=0.30\linewidth]{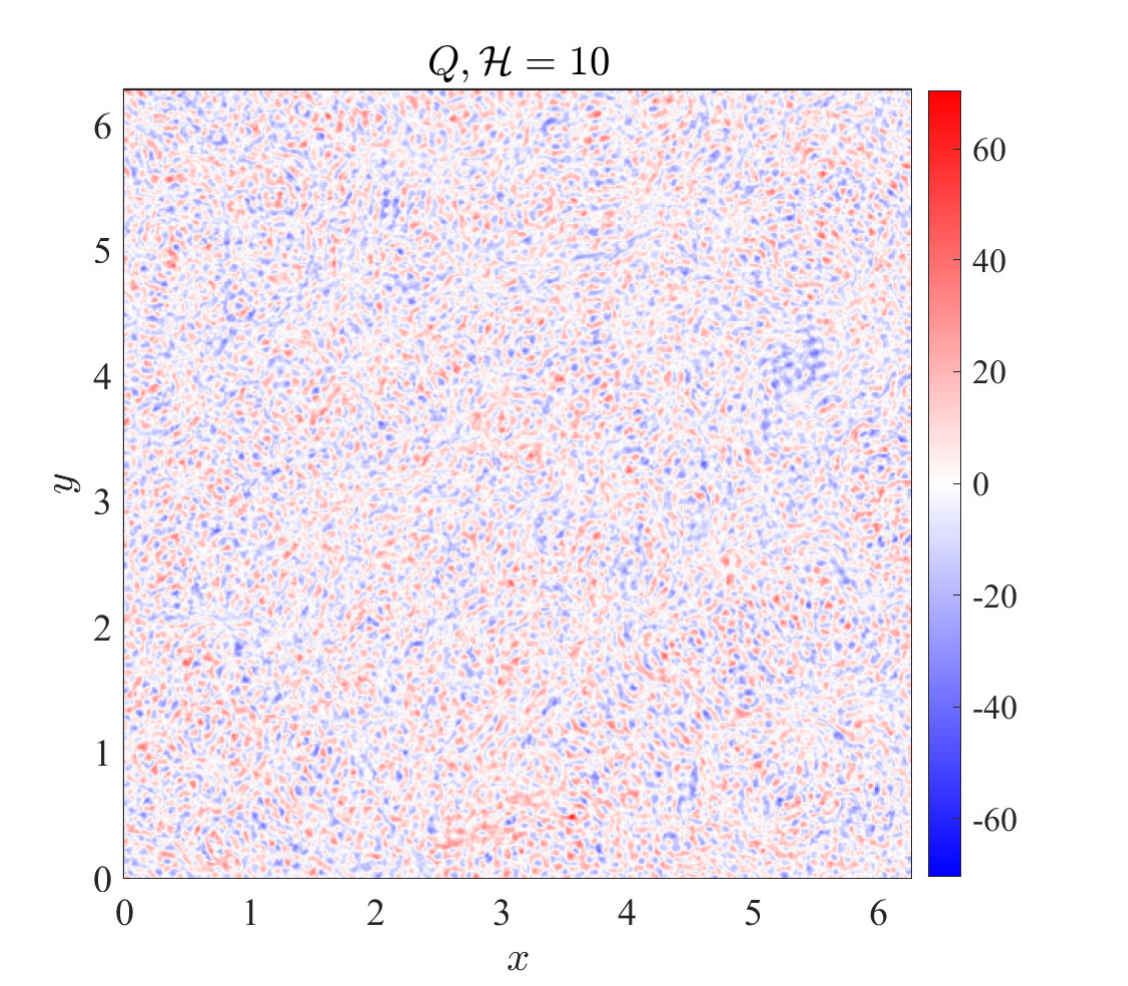}
    \includegraphics[width=0.30\linewidth]{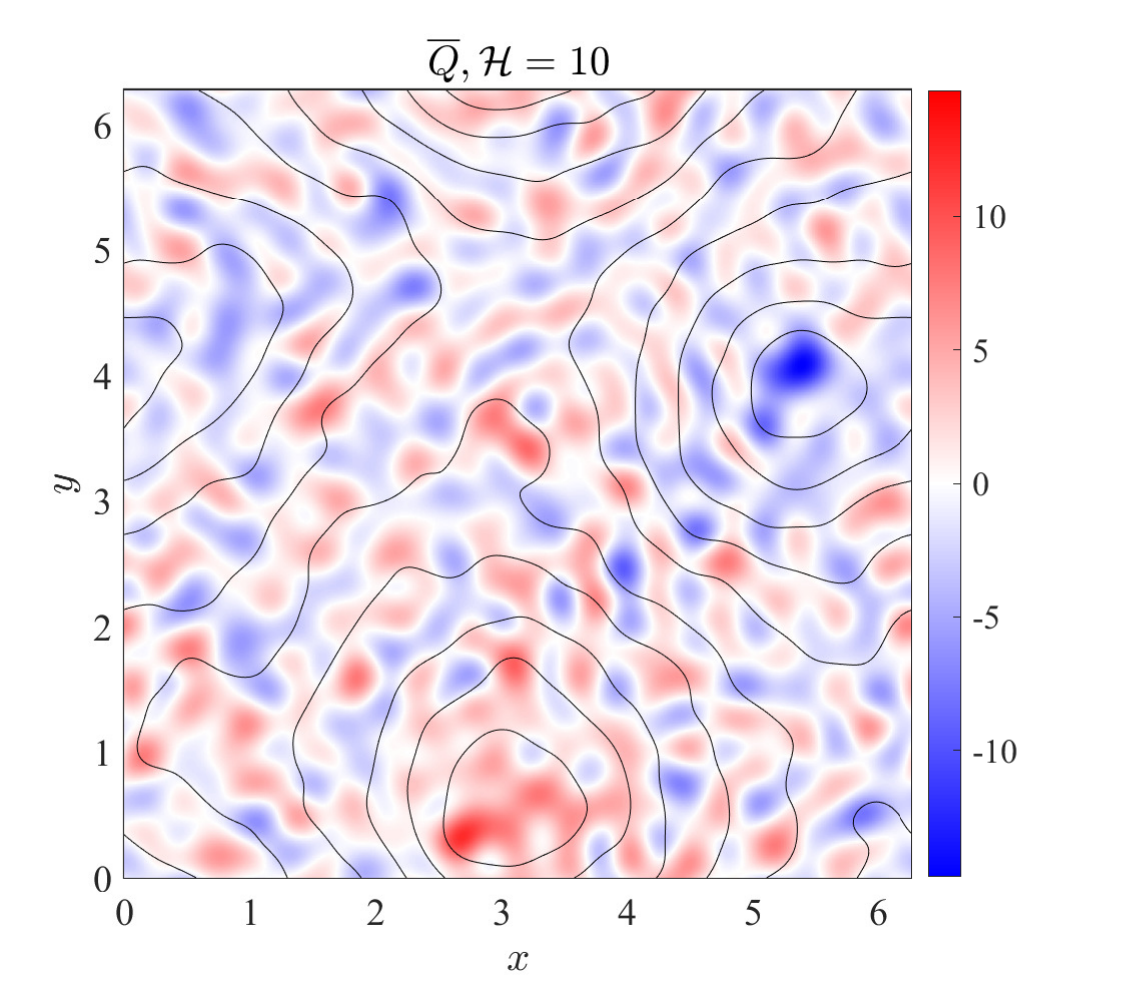}
    \includegraphics[width=0.30\linewidth]{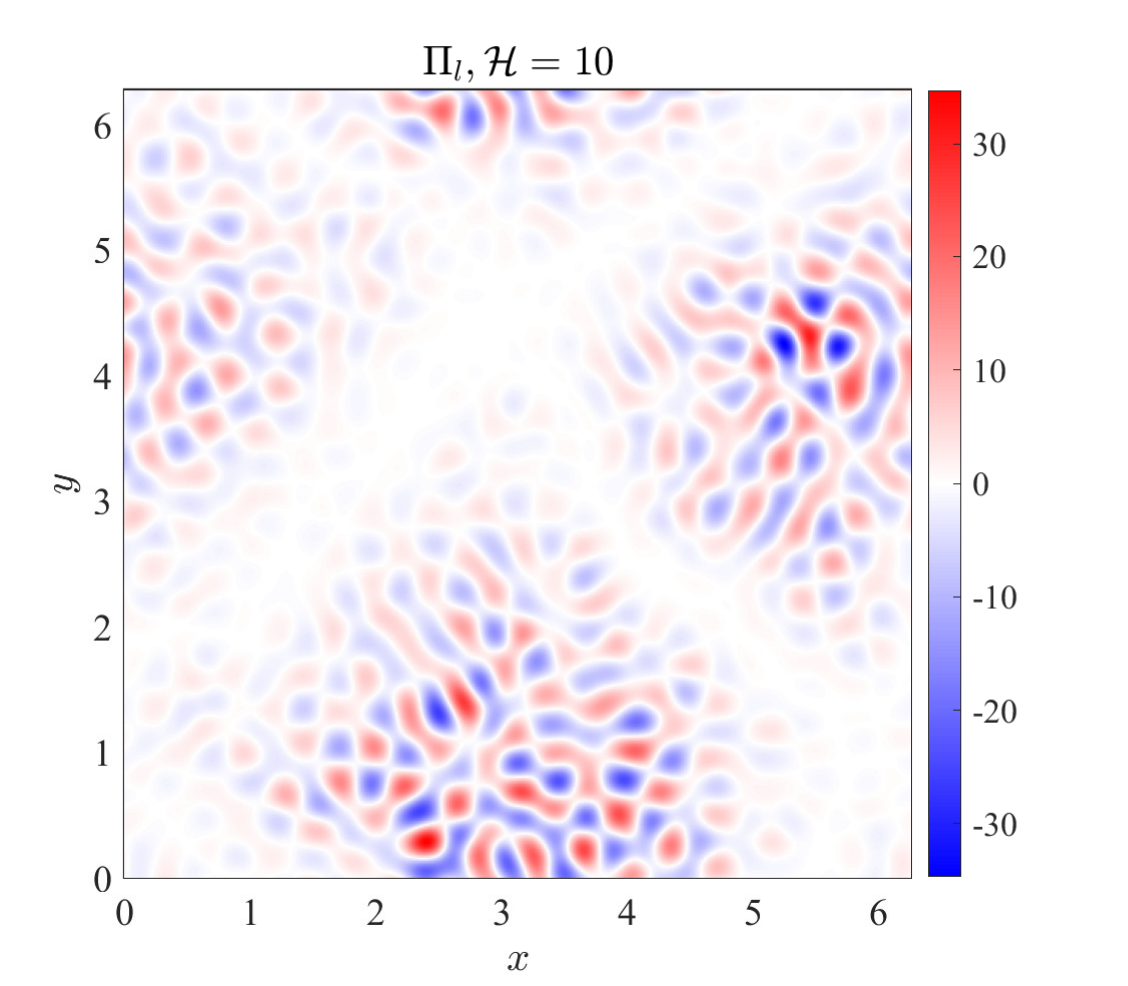}
	\caption{The snapshots of $Q$, $\overline{Q}$, $\Pi_l$ at $\mathcal{H}=10$ with $\epsilon=0.008,k_h=64$. The black curves in the snapshots of $\bar{Q}$ are the contours of the filtered stream-function $\bar{\psi}$.
	Ten contours range in $[-1.60,1.56]$.
	}
	\label{fig:Mh32SnapShots}
\end{figure}

\begin{figure}
	\centering
	\includegraphics[width=0.30\linewidth]{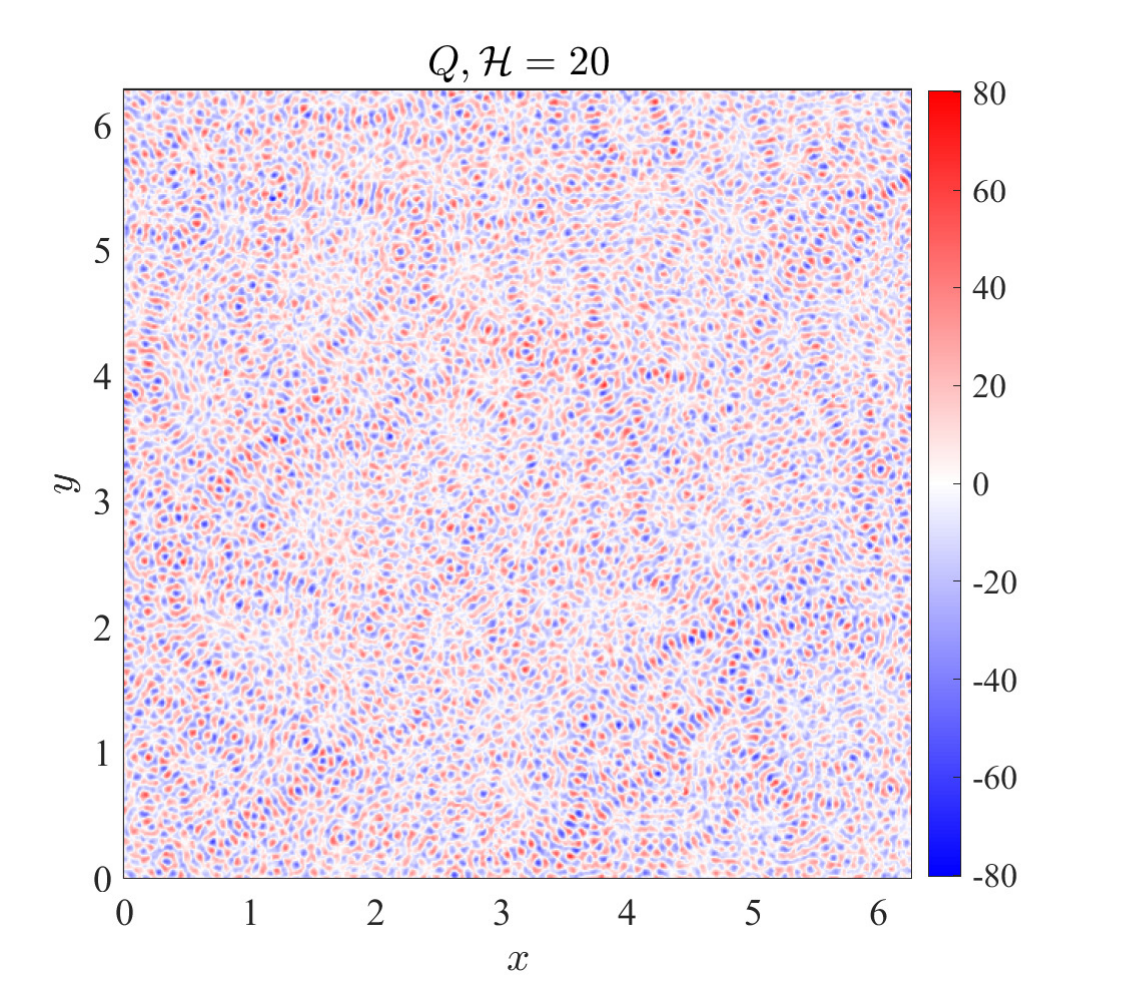}
    \includegraphics[width=0.30\linewidth]{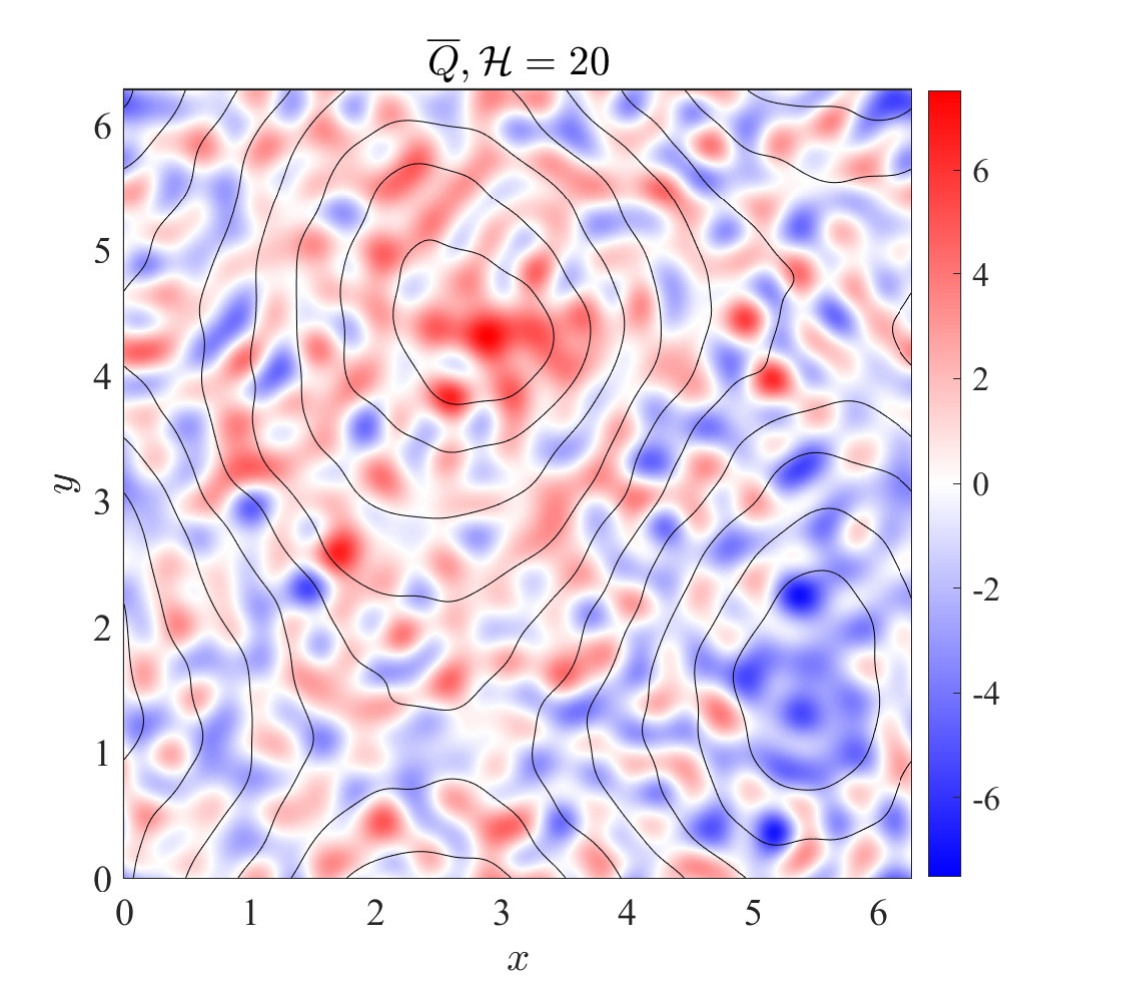}
    \includegraphics[width=0.30\linewidth]{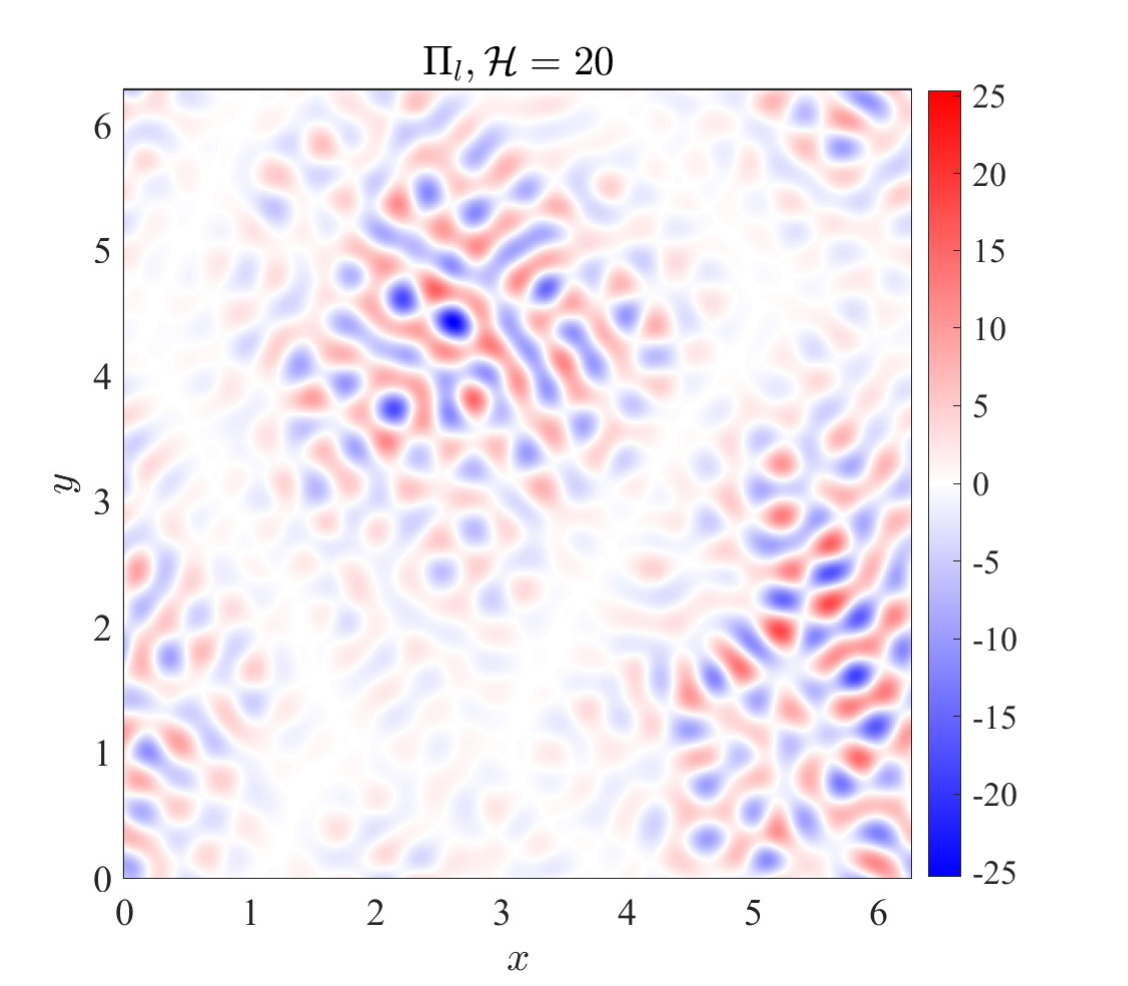}
	\caption{The snapshots of $Q$, $\overline{Q}$, $\Pi_l$ at $\mathcal{H}=20$ with $\epsilon=0.008,k_h=64$. The black curves in the snapshots of $\bar{Q}$ are the contours of the filtered stream-function $\bar{\psi}$.
	Ten contours range in $[-1.50,1.42]$.
	}
	\label{fig:Mh64SnapShots}
\end{figure}

\begin{figure}
	\centering
	\includegraphics[width=0.30\linewidth]{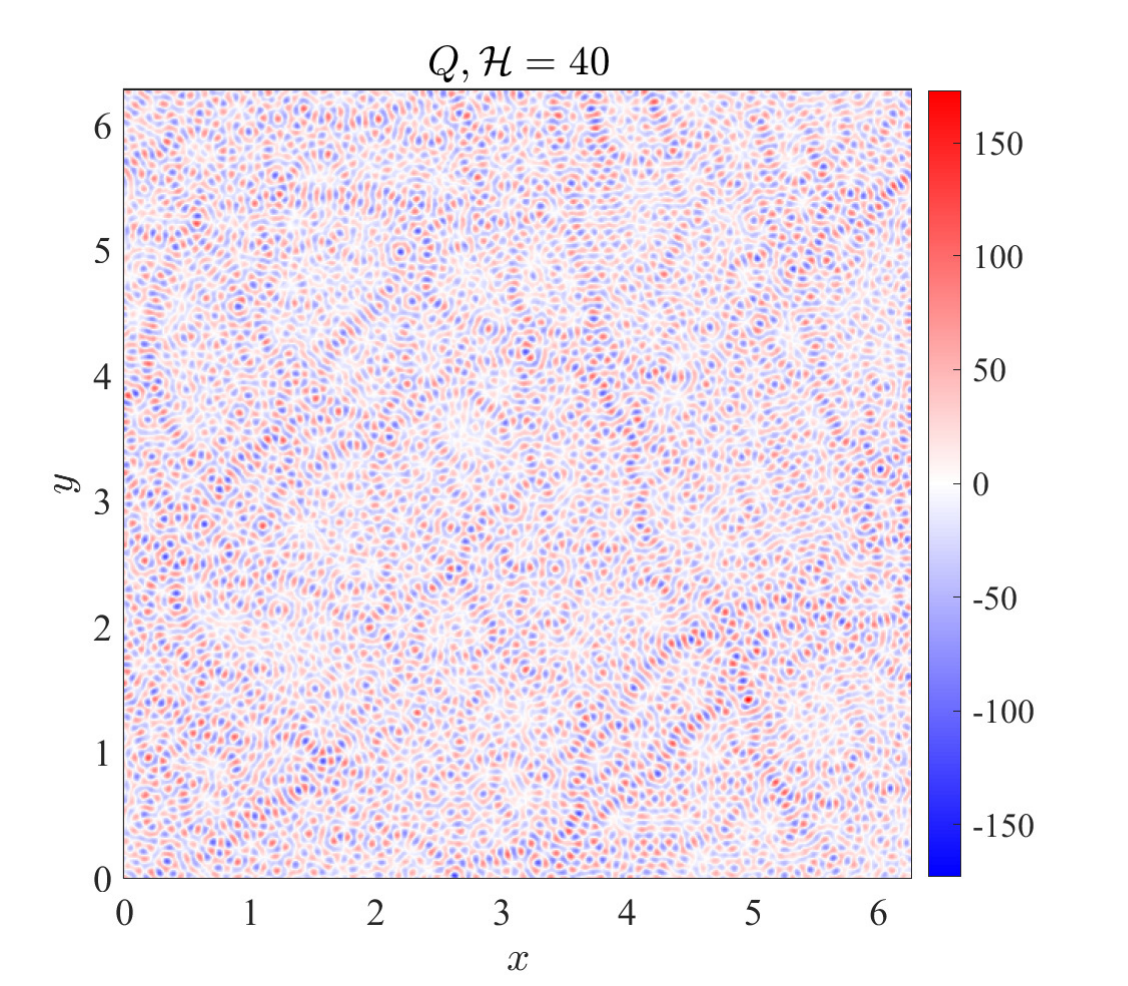}
    \includegraphics[width=0.30\linewidth]{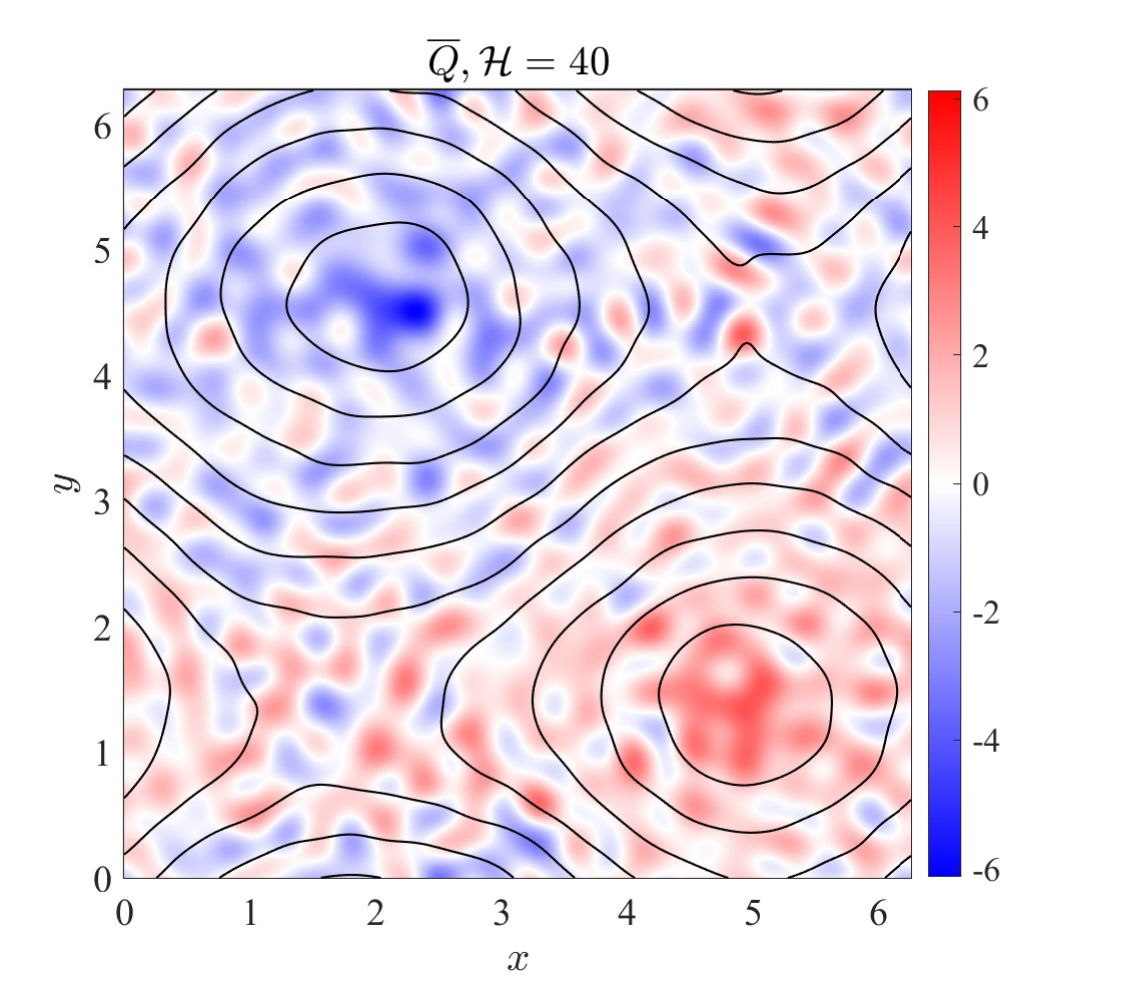}
    \includegraphics[width=0.30\linewidth]{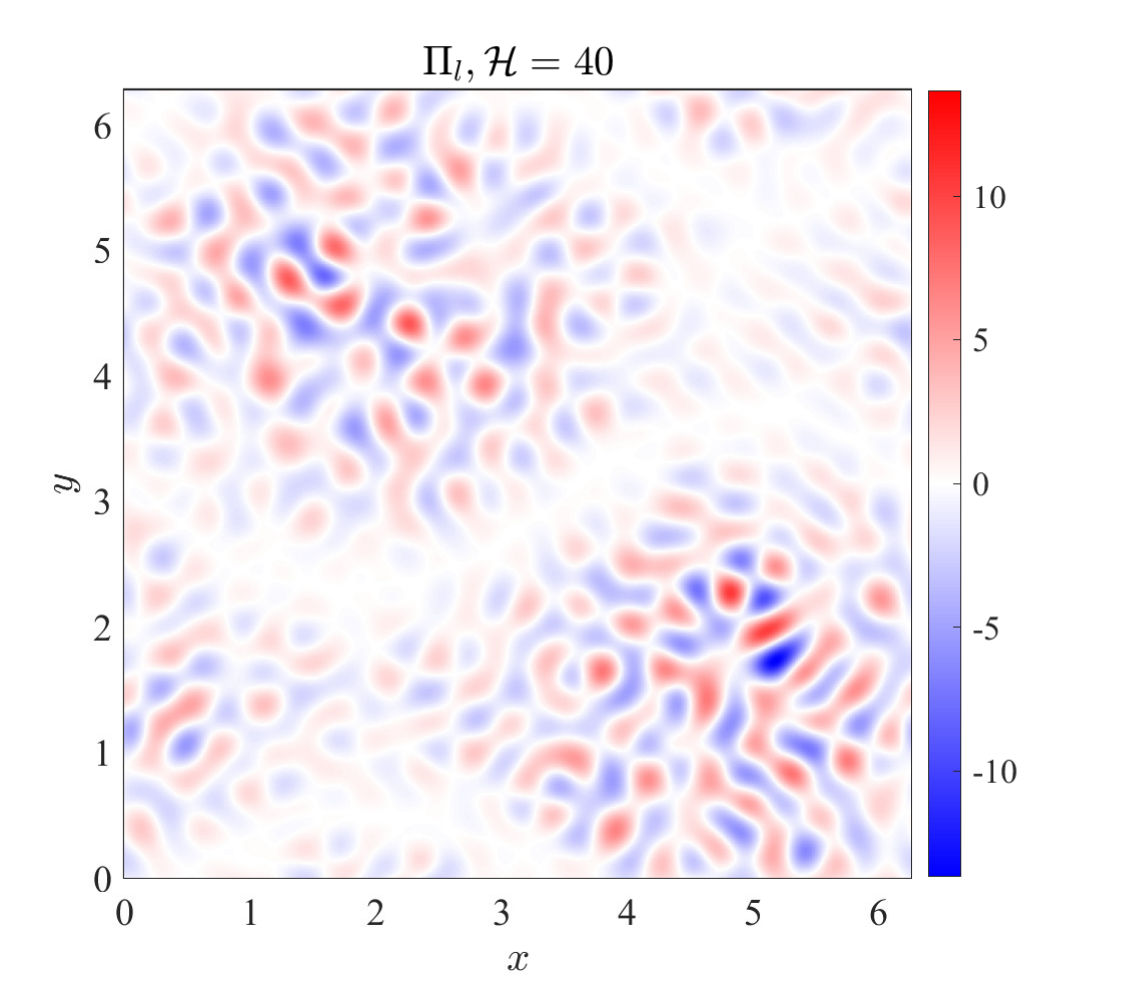}	
	\caption{The snapshots of $Q$, $\overline{Q}$, $\Pi_l$ at $\mathcal{H}=40$ with $\epsilon=0.008,k_h=64$. The black curves in the snapshots of $\bar{Q}$ are the contours of the filtered stream-function $\bar{\psi}$.
	Ten contours range in $[-1.31,1.33]$.
	}
	\label{fig:Mh128SnapShots}
\end{figure}

\begin{figure}
	\centering
	\includegraphics[width=0.30\linewidth]{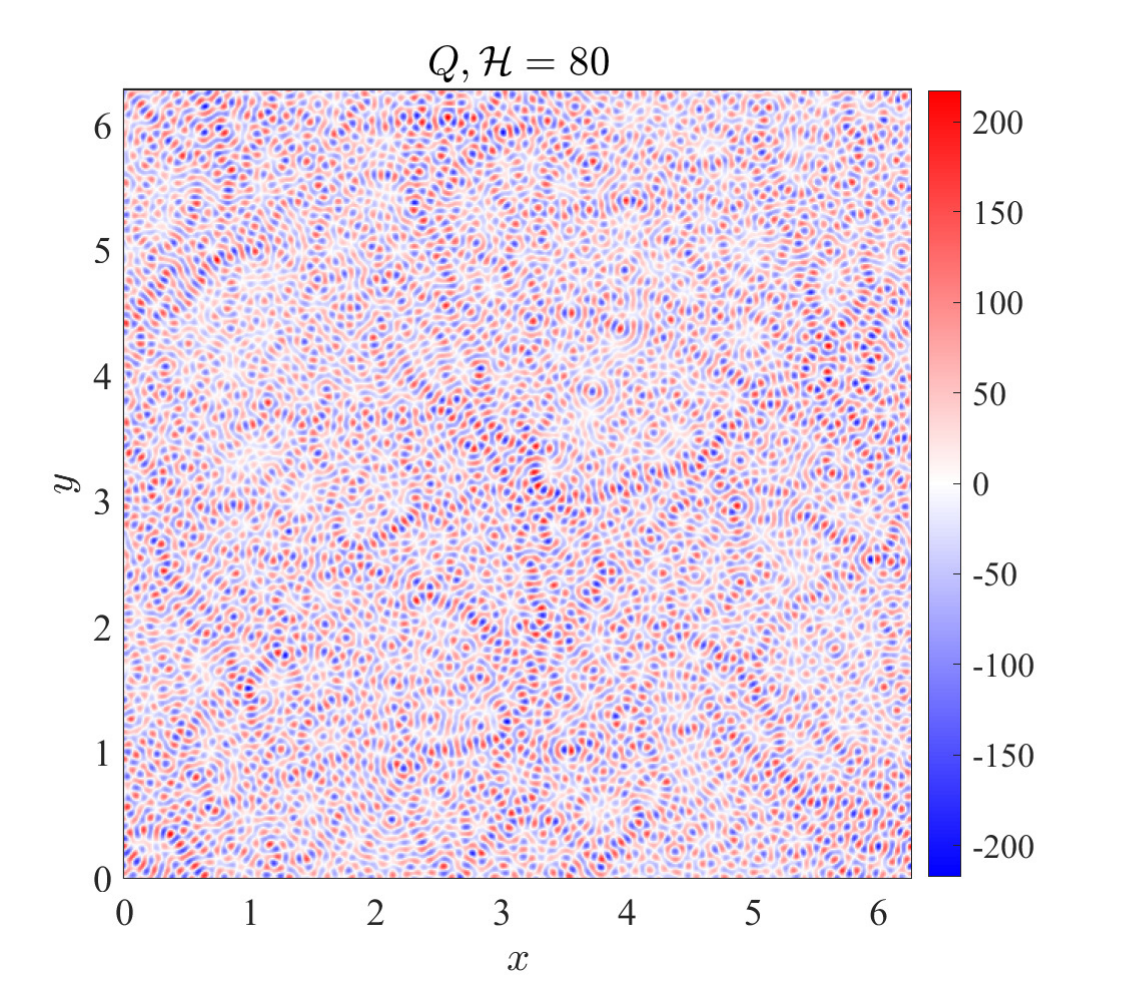}
    \includegraphics[width=0.30\linewidth]{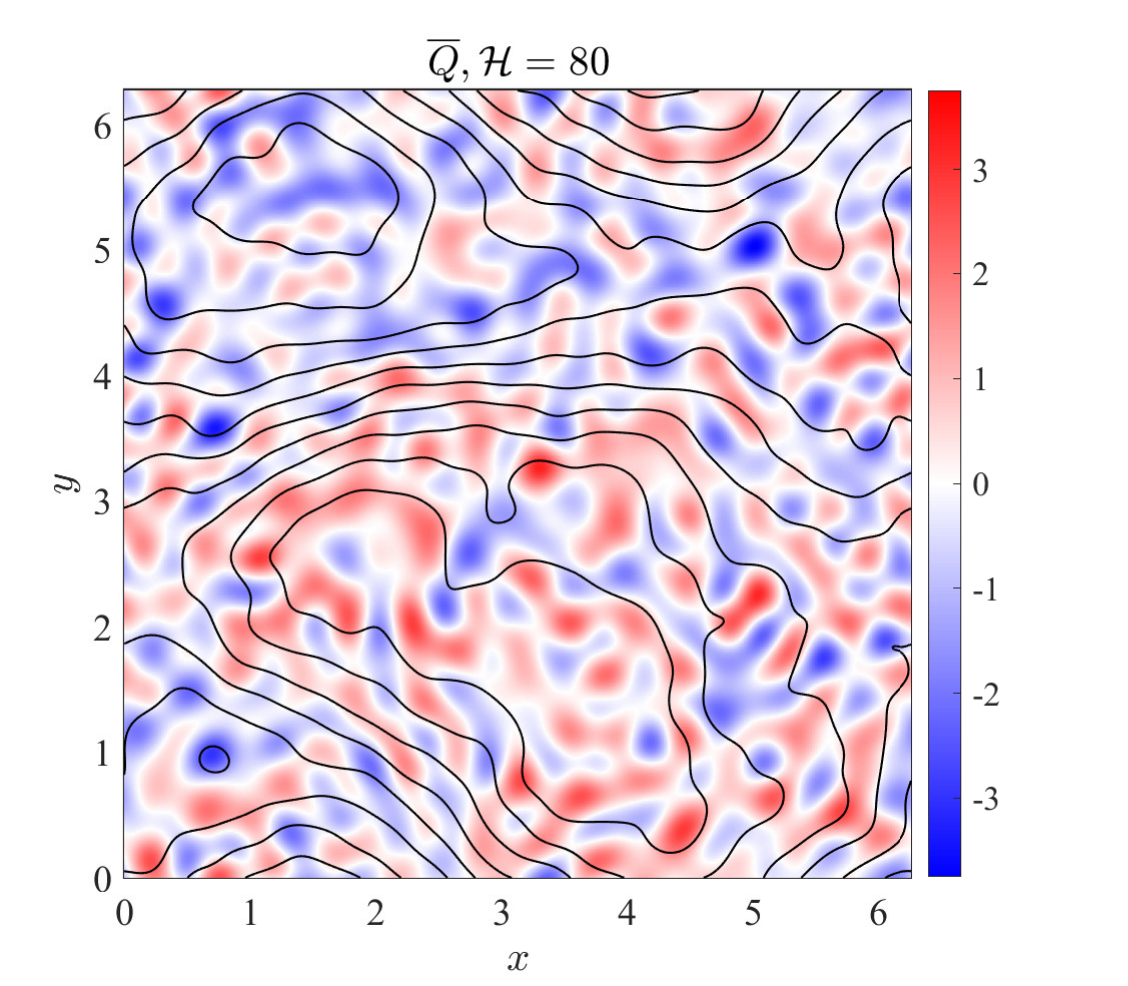}
    \includegraphics[width=0.30\linewidth]{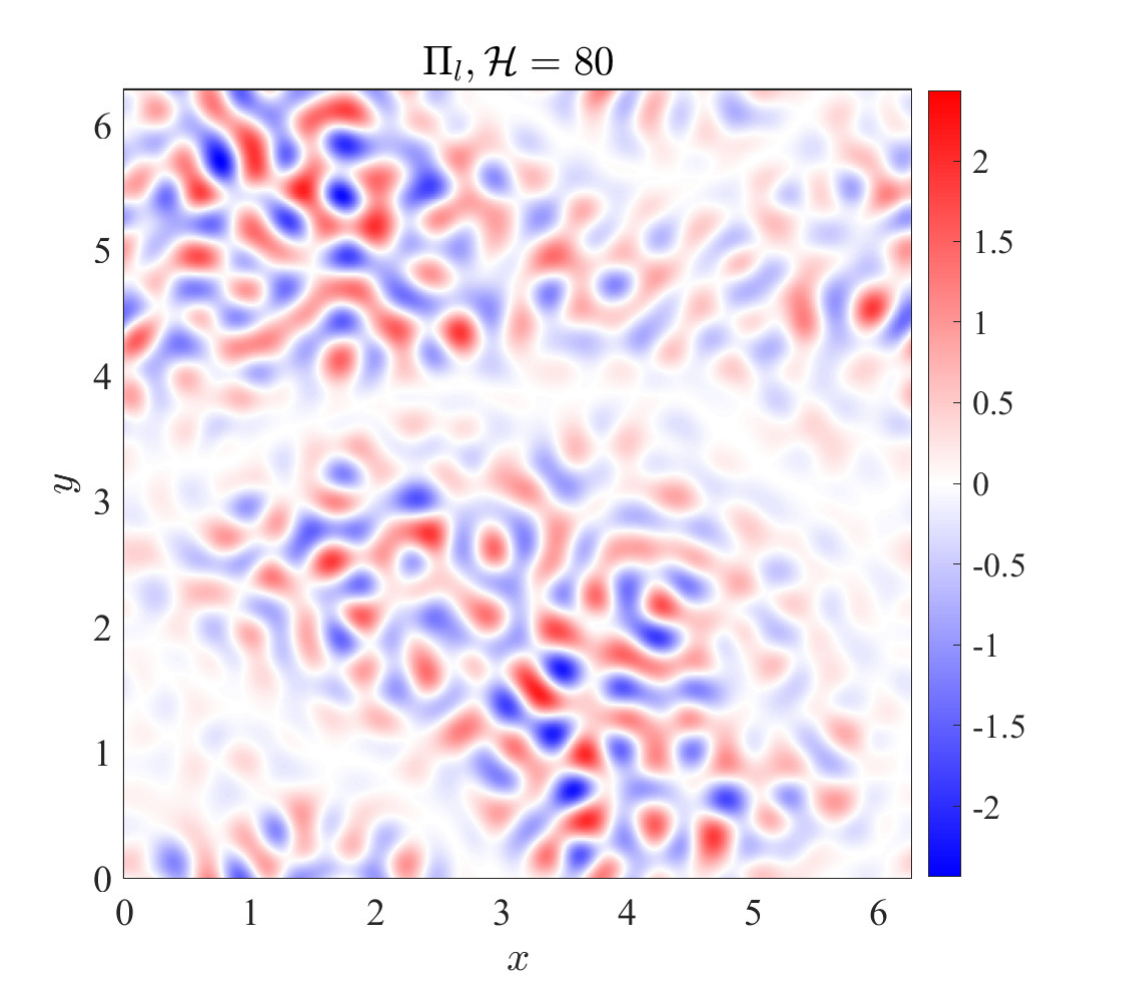}
	\caption{The snapshots of $Q$, $\overline{Q}$, $\Pi_l$ at $\mathcal{H}=80$ with $\epsilon=0.008,k_h=64$. The black curves in the snapshots of $\bar{Q}$ are the contours of the filtered stream-function $\bar{\psi}$.
	Ten contours range in $[-0.27,0.39]$.
	}
	\label{fig:Mh256SnapShots}
\end{figure}

\begin{figure}
	\centering
	\includegraphics[width=0.30\linewidth]{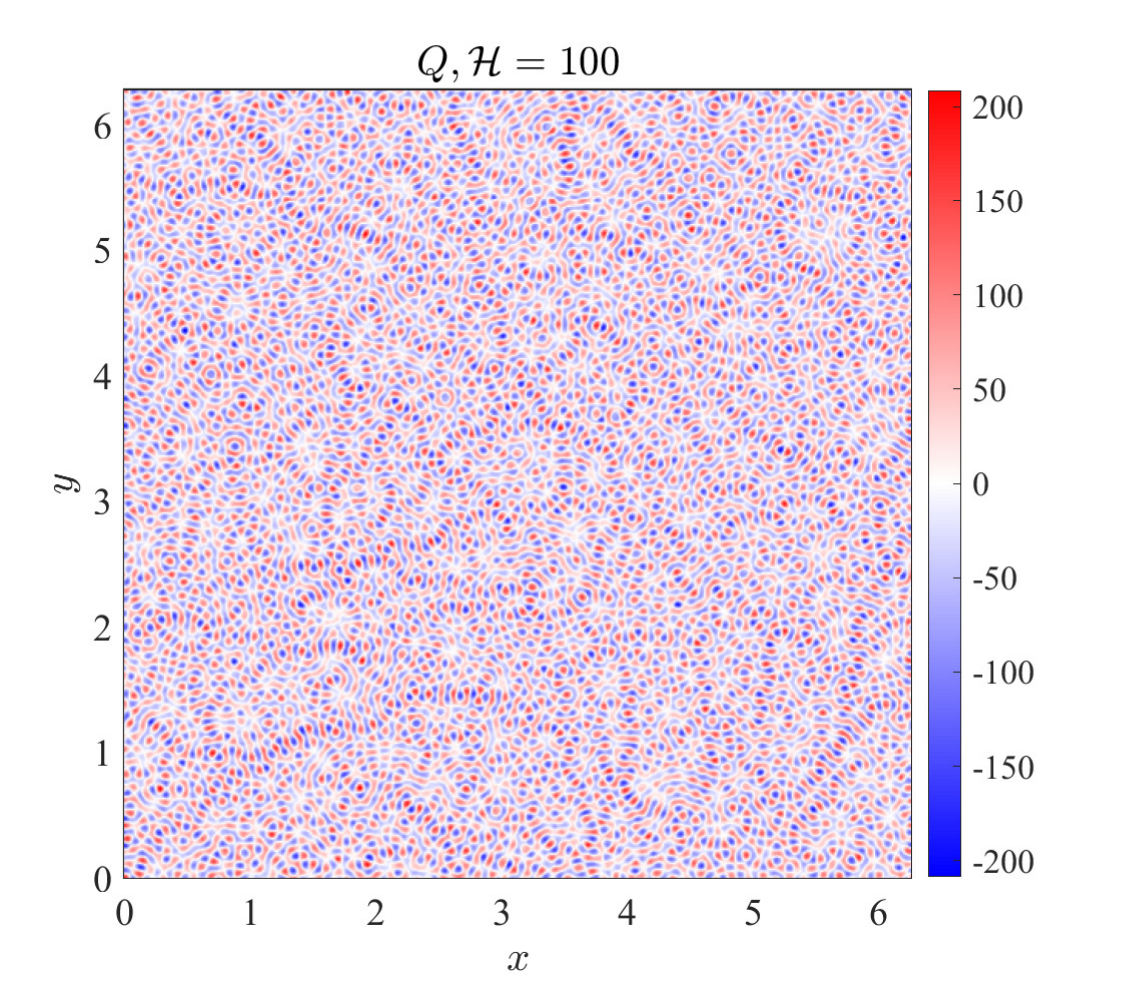}
    \includegraphics[width=0.30\linewidth]{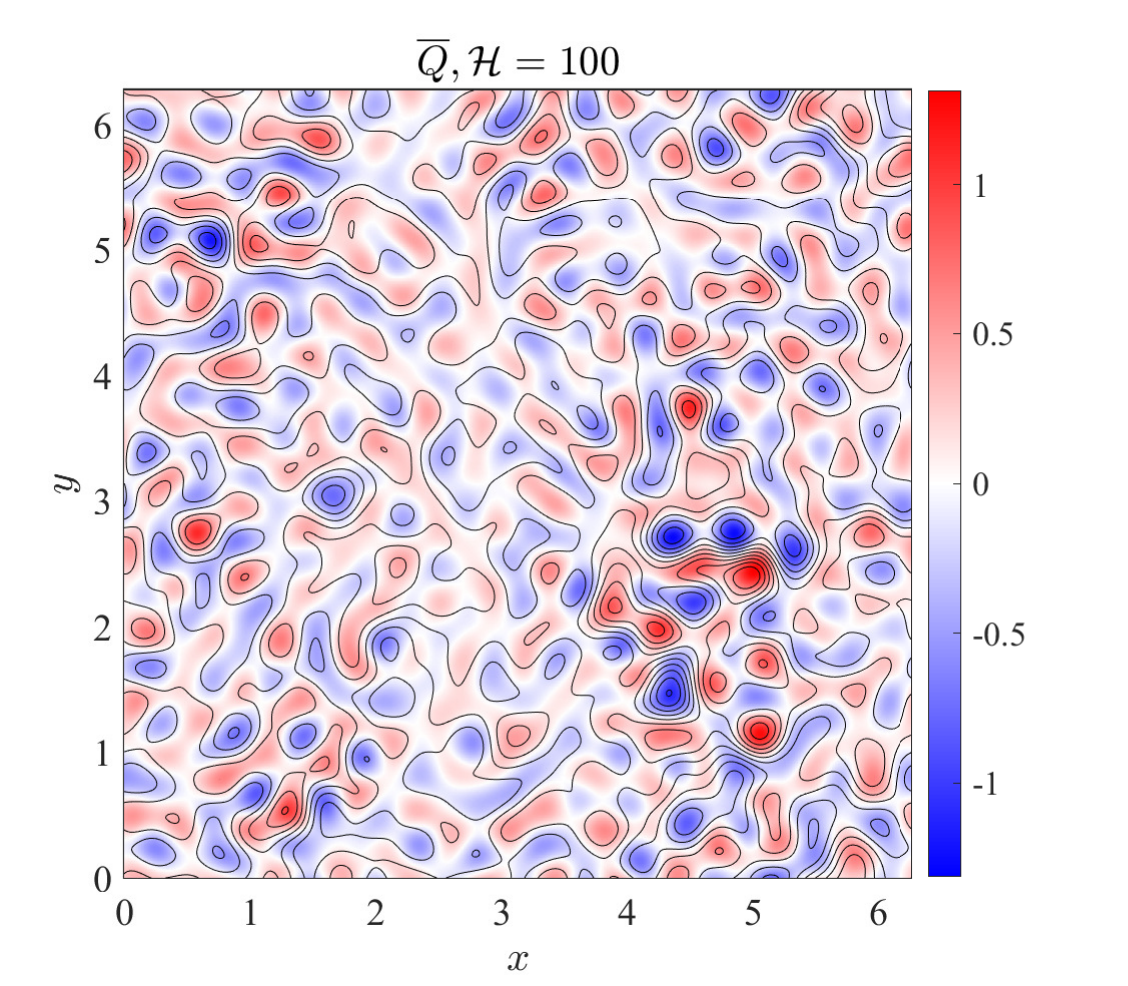}
    \includegraphics[width=0.30\linewidth]{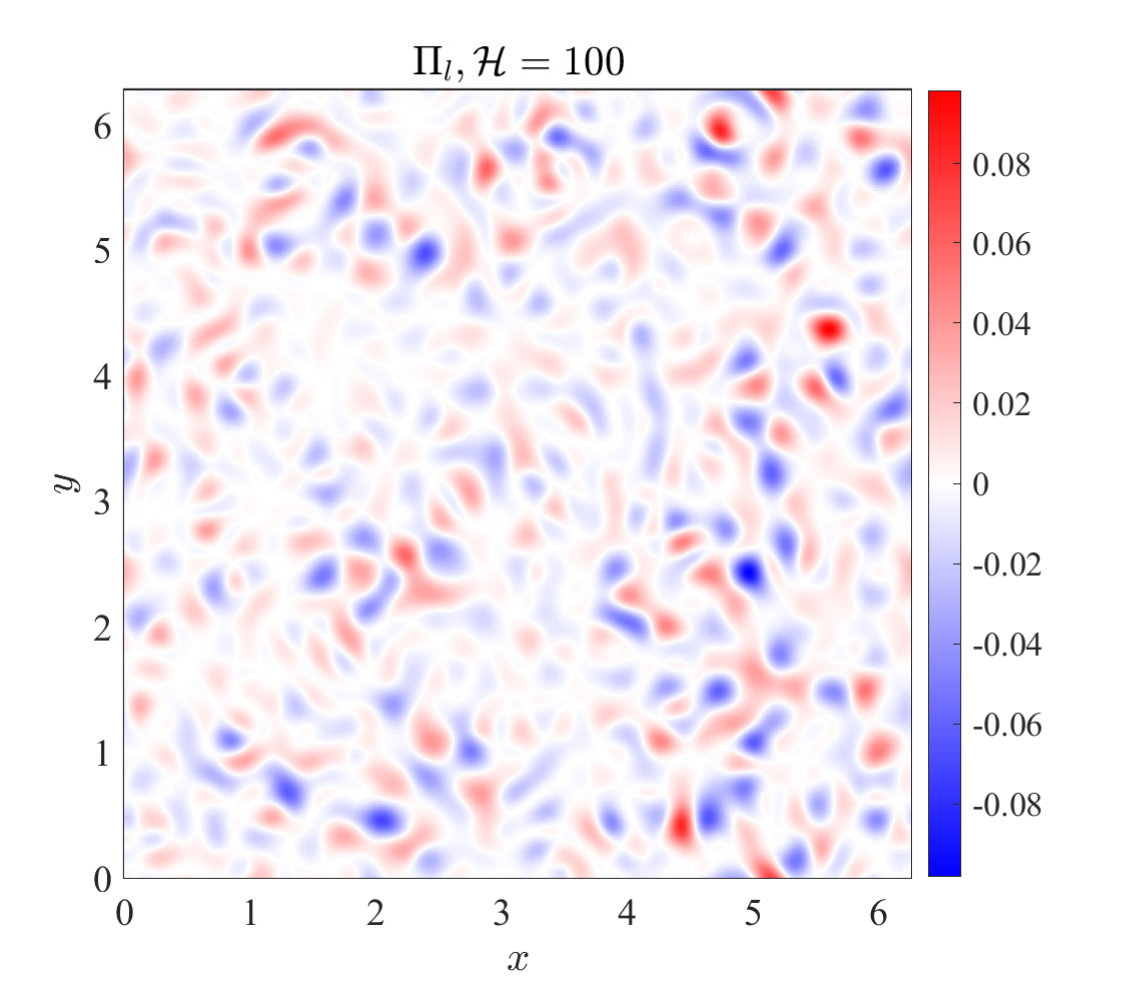}
	\caption{The snapshots of $Q$, $\overline{Q}$, $\Pi_l$ at $\mathcal{H}=100$ with $\epsilon=0.008,k_h=64$. The black curves in the snapshots of $\bar{Q}$ are the contours of the filtered stream-function $\bar{\psi}$.
	Ten contours range in $[-0.01,0.01]$.
	}
	\label{fig:Mh320SnapShots}
\end{figure}

According to Radko's multi-scale method\cite{radko_spin_down_2022a}, the leading order balance is given by $\nabla^2\psi_1 = -h$. 
Consequently, as $\mathcal{H}$ increases (as shown in Figures~\ref{fig:Mh16SnapShots} to \ref{fig:Mh320SnapShots}), the $Q$ fields gradually become dominated by topography. 
However, the filtered vorticity fields $\overline{Q}$ within the range $\mathcal{H}\in[10,80]$ exhibit large-scale structures, indicating energy condensation at the domain size. 
The magnitude of $\Pi_l$ is higher in the regions with these large-scale structures, further reflecting energy condensation.

When $\mathcal{H}=100$, energy condensation disappears, and the magnitude of $\Pi_l$ approaches zero.

\section{Characteristic length scale}\label{scale}
\begin{figure}
	\centering
	\includegraphics[width=0.65\linewidth]{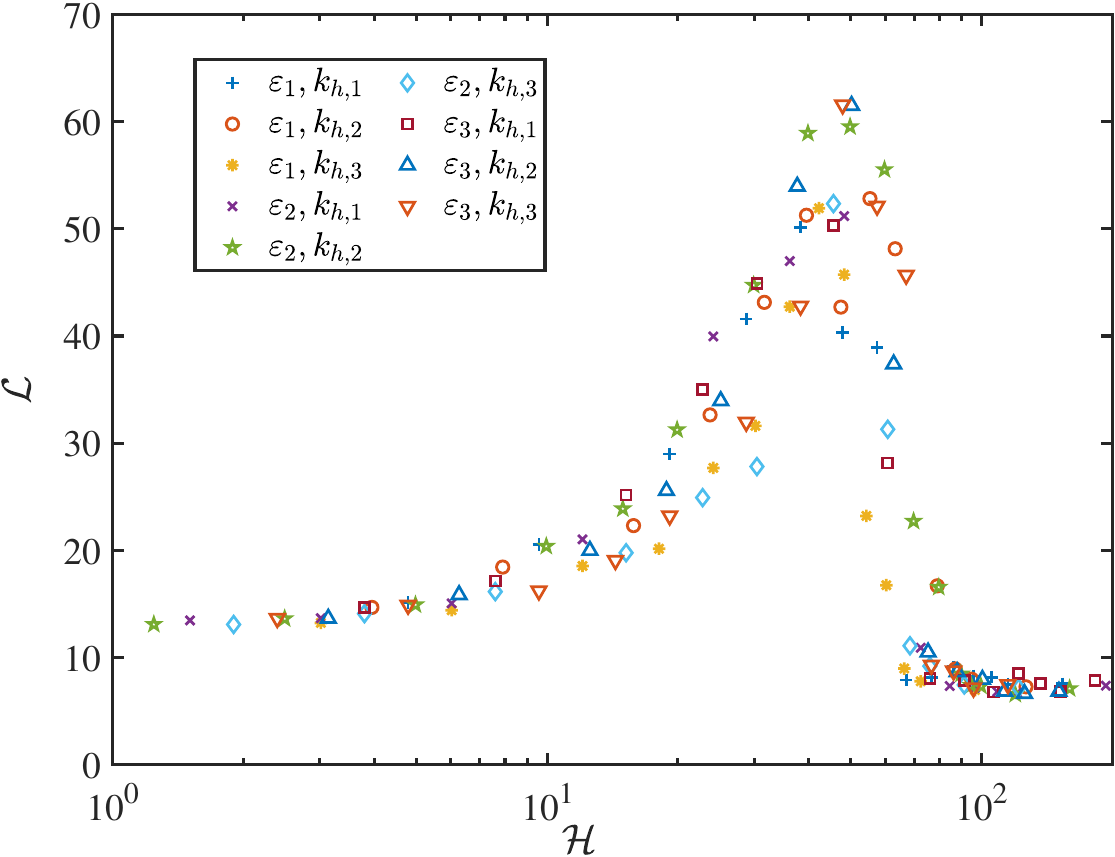}
	\caption{The dependence of the characteristic length scale on the magnitude of topography $\mathcal{H}$ with $\epsilon=0.008, k_h=64$.}
	\label{fig:CorrScale}
\end{figure}

In this section, we use the characteristic length scale to quantify the energy condensation, which reflects the coherence of filtered vorticity $\overline{Q}$.
We calculate the vorticity correlation function as 
\begin{equation}
    	C_{Q}(\boldsymbol{r})=\langle \overline{Q}(\boldsymbol{x}+\boldsymbol{r})\overline{Q}(\boldsymbol{x}) \rangle
\end{equation}
The characteristic length scale, denoted as $\mathcal{L}$, is then obtained as:
\begin{equation}
    	\mathcal{L}\equiv\frac{1}{ C_{Q}(0)}\int_0^\infty C_{Q}(r) \mdif r
\end{equation}
Figure~\ref{fig:CorrScale} illustrates the dependence of the characteristic length scale on the normalized magnitude of topography $\mathcal{H}$.
We observe that $\mathcal{L}$ increases to maximum at $\mathcal{H} \approx 50$.
For $\mathcal{H}>50$, $\mathcal{L}$ decreases, reaching its minimum as $\mathcal{H}$ approaches the critical point $\mathcal{H}_c\approx 90$.
This critical point corresponds to the second-order transition observed in the dependence of energy transfer $T(K=1)$ on $\mathcal{H}$ (cf. Figure 2(c) in the letter).
The peak location of $\mathcal{L}$ in figure~\ref{fig:CorrScale} is different from that of $T(K=1)$ at $\mathcal{H} \approx 20$.
Since the local energy transfer weankens and nonlocal energy transfer strengthens as we present in the next section, the energy condensed at the domain size decreases less than that at the energy inertial range (cf. fig 2 (a) in the letter), which enhances the characteristic length in the range of $\mathcal{H}\in(20,50)$.

%\section{Energy flux}

\section{Energy transfer cross scale}\label{transfer}
\begin{figure}[H]
	\centering
	\includegraphics[width=0.30\linewidth]{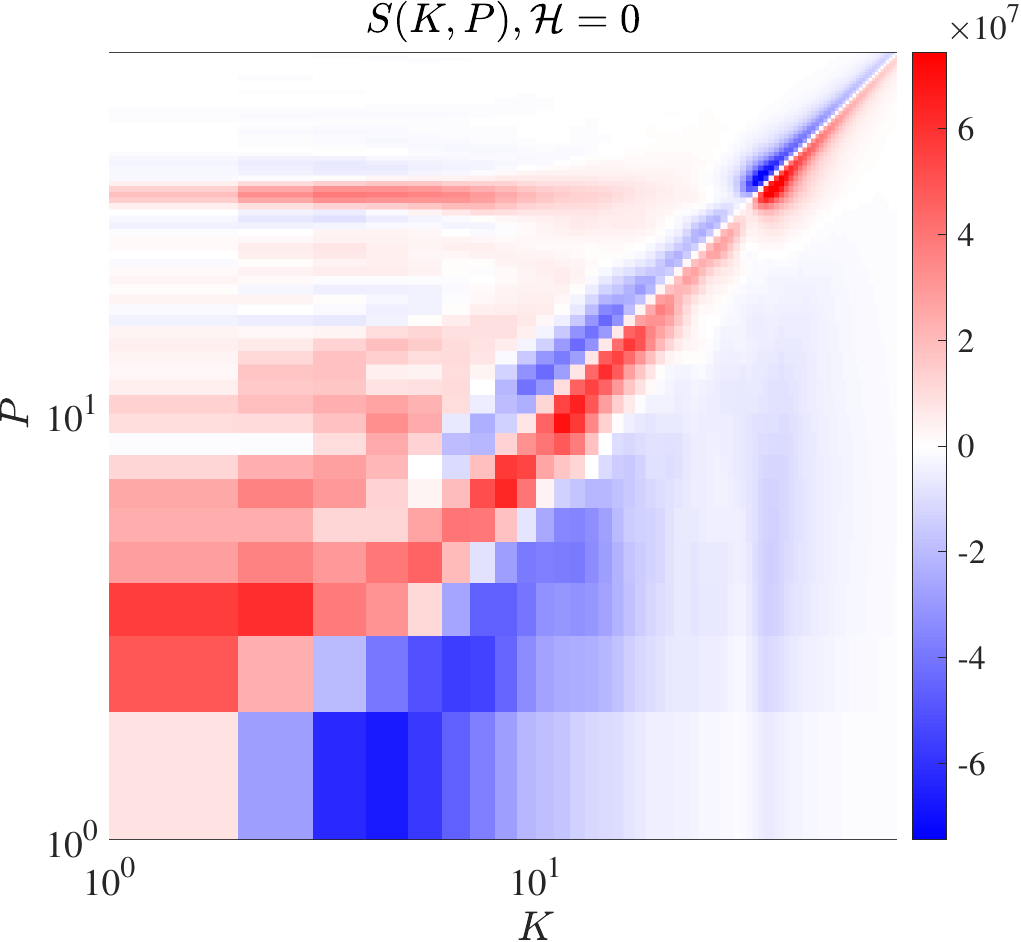}
	\includegraphics[width=0.30\linewidth]{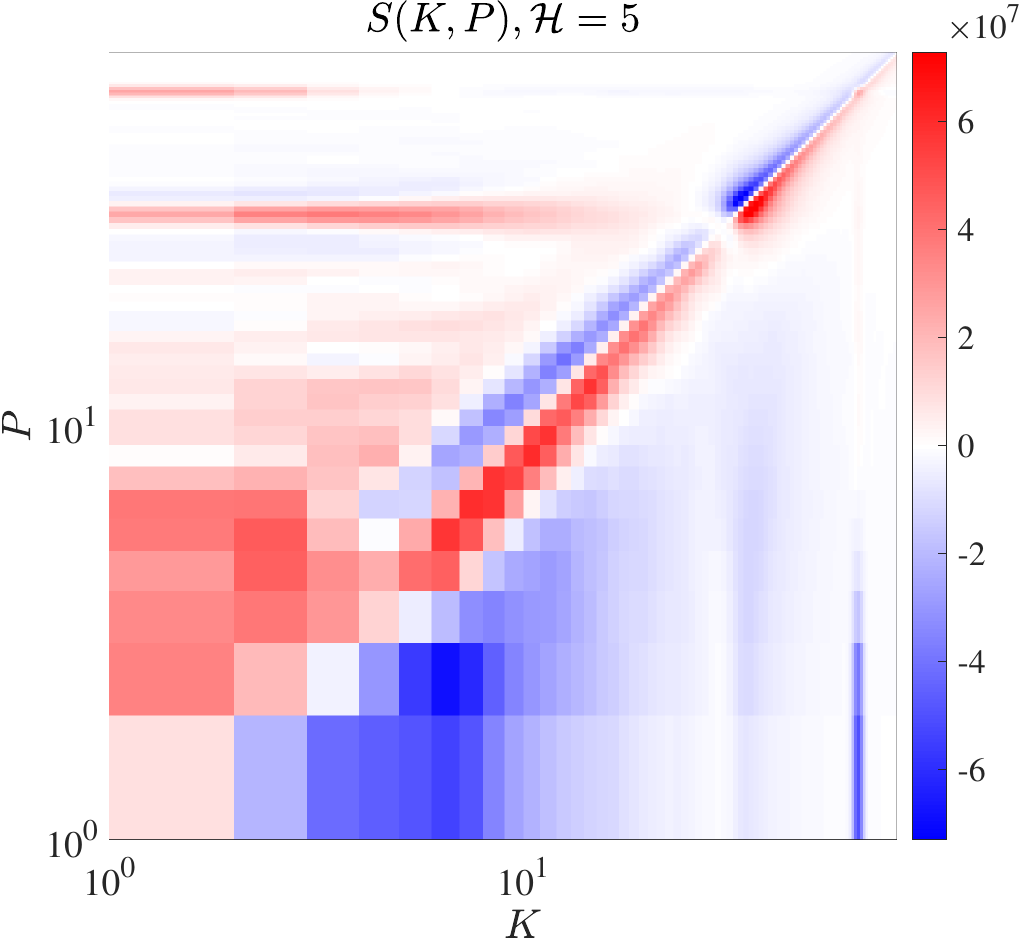}
	\includegraphics[width=0.30\linewidth]{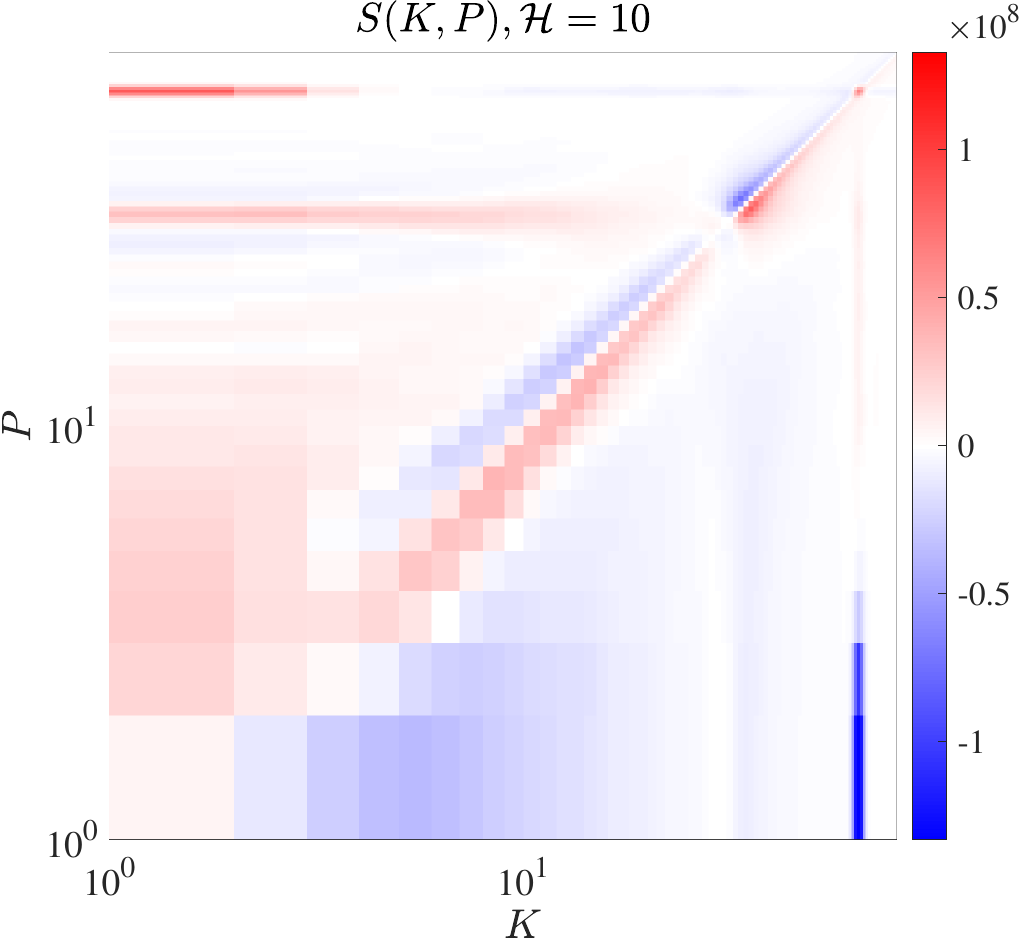}
	\includegraphics[width=0.30\linewidth]{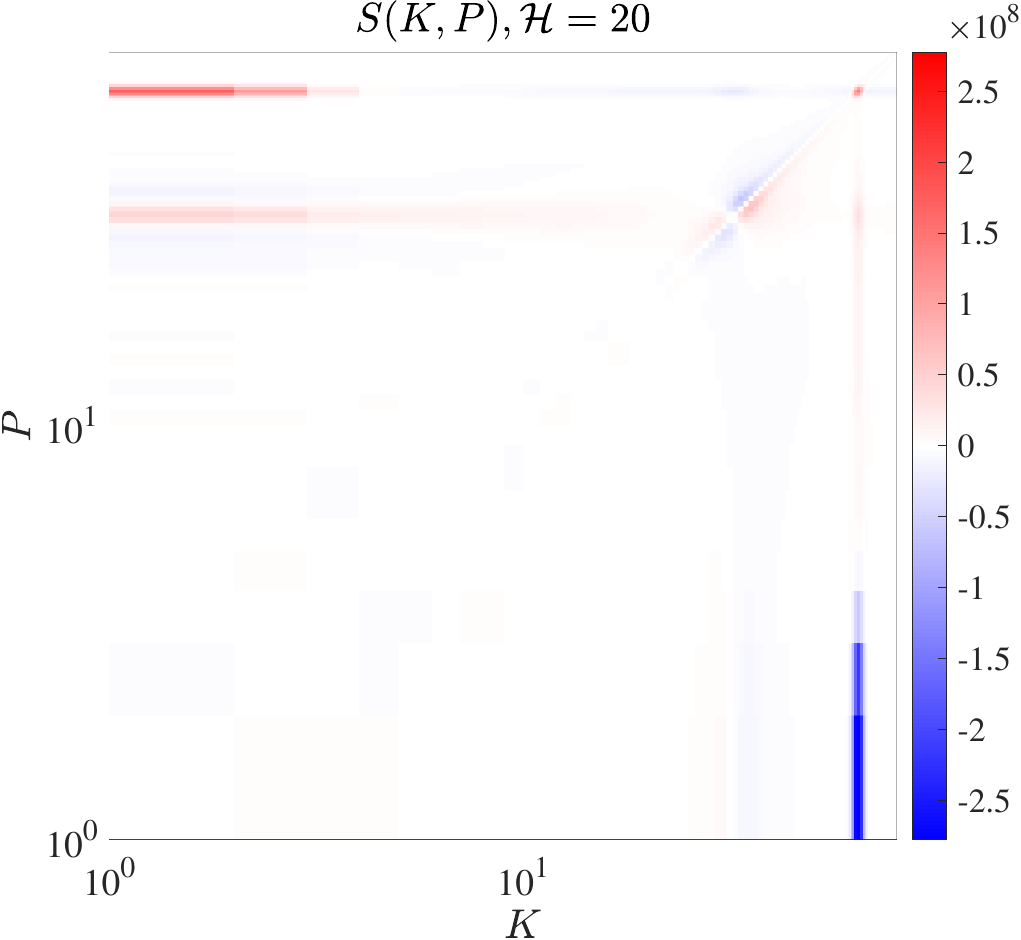}
    \includegraphics[width=0.30\linewidth]{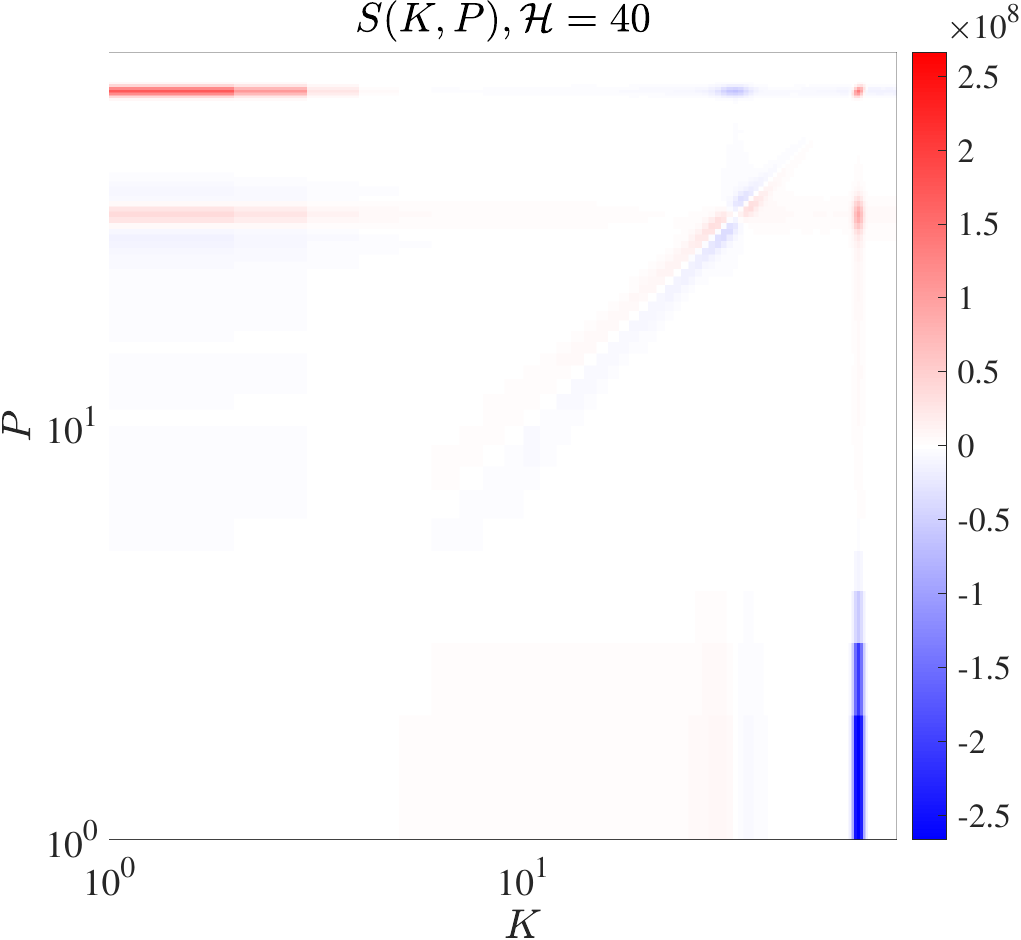}
	\includegraphics[width=0.30\linewidth]{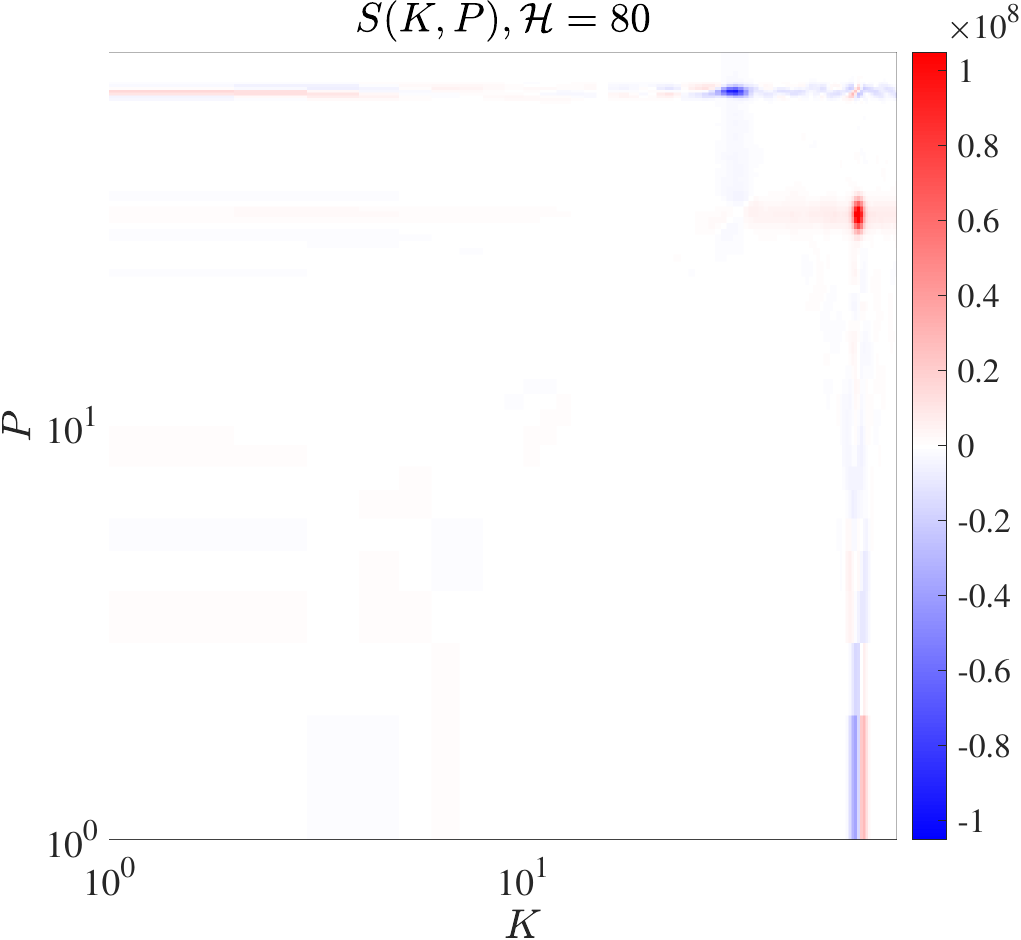}
	\includegraphics[width=0.30\linewidth]{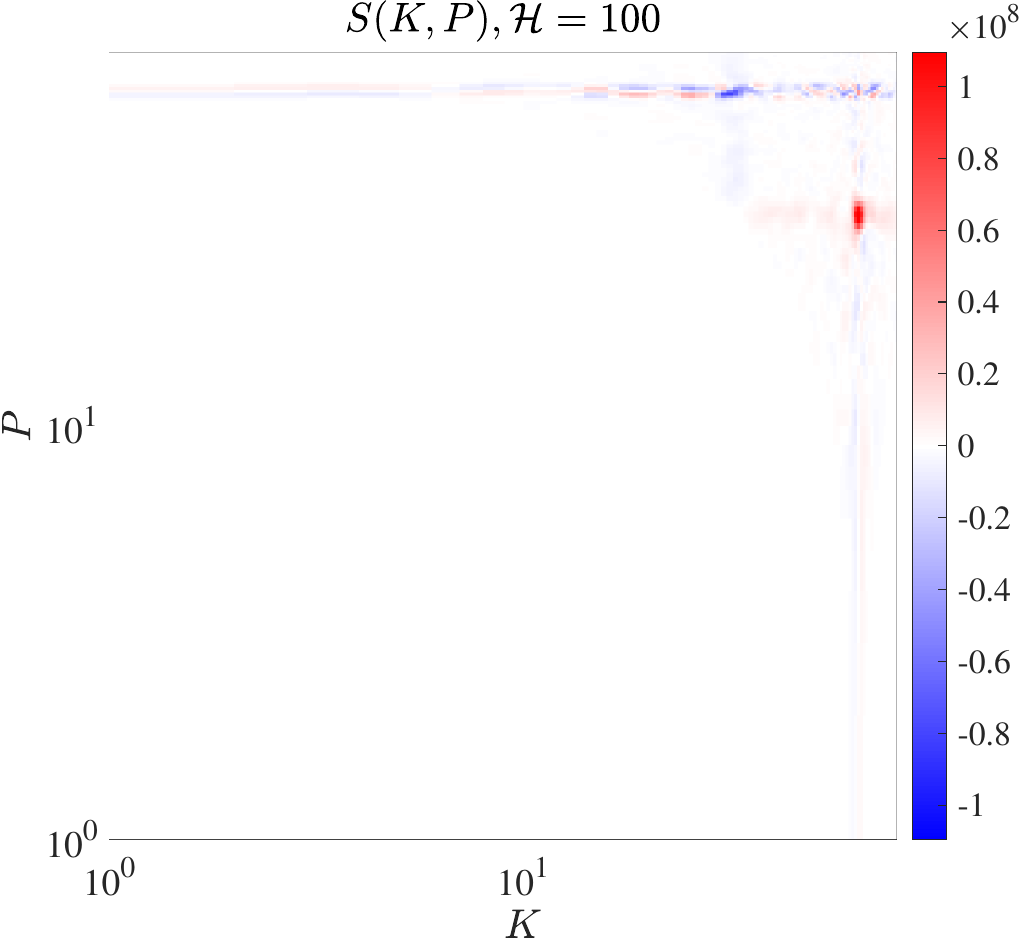}
	\caption{The shell to shell energy transfer with $\epsilon=0.008$ and $k_h=64$.}
	\label{fig:EnergyNonLinearMh}
\end{figure}

The interaction in the nonlinear term involves a wave number traid $(\boldsymbol{k},\boldsymbol{p},\boldsymbol{r})$ satisfying $\boldsymbol{k}+ \boldsymbol{p}+\boldsymbol{r} = 0$.
Here $\boldsymbol{k}=(k_x,k_y)$, $\boldsymbol{p}=(p_x,p_y)$ and $\boldsymbol{r}=(r_x,r_y)$.
In a traid, the combined energy transfer from the mode $\boldsymbol{p}$ and $\boldsymbol{r}$ is equal to the energy gained by the mode $\boldsymbol{k}$ \cite{kraichnan_inertial_range_1971,lesieur_turbulence_2008}.
In our system, considering the energy transfer in three mode traid, the combined energy transfer from mode $\boldsymbol{p}$ and $\boldsymbol{r}$ to mode $\boldsymbol{k}$ is 
\begin{equation}
	S(\boldsymbol{k}|\boldsymbol{p},\boldsymbol{r}) =  \frac{1}{2}\left((p_x r_y-p_y r_x)\mathsf{Re}[\hat{\psi}_{\boldsymbol{k}} \hat{\psi}_{\boldsymbol{p}} \hat{q}_{\boldsymbol{r}}] + (r_x p_y-r_y p_x)\mathsf{Re}[\hat{\psi}_{\boldsymbol{k}} \hat{\psi}_{\boldsymbol{r}} \hat{q}_{\boldsymbol{p}}] \right)\delta(\boldsymbol{k}+\boldsymbol{p}+\boldsymbol{r}).
\end{equation}
Here, $q=Q+h$ denotes the potential vorticity.
The combined energy transferred to mode $\boldsymbol{p}$ and $\boldsymbol{r}$ is similar to that for $\boldsymbol{k}$.
After some algebra, we have
\begin{equation}
	S(\boldsymbol{k}|\boldsymbol{p},\boldsymbol{r}) + S(\boldsymbol{p}|\boldsymbol{k},\boldsymbol{r})+S(\boldsymbol{r}|\boldsymbol{k},\boldsymbol{p}) = 0.
\end{equation}
By summing over $\boldsymbol{r}$ and summing $\boldsymbol{k}$ in the shell $K$ and $\boldsymbol{p}$ in the shell $P$, we obtain the corresponding energy transfer (cf. \cite{rubio_upscale_2014,plumley_effects_2016}) as
\begin{equation}
    	S(K,P) = \sum_{|\boldsymbol{k}|=K} \sum_{|\boldsymbol{p}|=P}\sum_{\boldsymbol{r}} S(\boldsymbol{k}|\boldsymbol{p},\boldsymbol{r}).
\end{equation}
Figure~\ref{fig:EnergyNonLinearMh} illustrates the energy transfer between shell $K$ and shell $P$ as $\mathcal{H}$ varies.
For $\mathcal{H} =0$, local energy transfer takes precedence. However, as $\mathcal{H}$ increases, local interactions progressively wane in influence, making room for non-local interactions to assume dominance. It is noteworthy that topography facilitates energy transfer from both the forcing and topographic scales to the system scale through non-local interactions.

\bibliography{ref}

\end{document}